\documentclass[journal]{IEEEtran}
\newcommand{\cev}[1]{\reflectbox{\ensuremath{\vec{\reflectbox{\ensuremath{#1}}}}}}

\usepackage{subfigure}
\usepackage{bm}
\usepackage{stackengine}

\usepackage{hyperref}
\usepackage[all]{hypcap}        
\usepackage{xcolor}
\usepackage[linesnumbered,ruled,vlined]{algorithm2e}
\usepackage{algorithmic}

\SetCommentSty{mycommfont}
\SetKwInput{KwInput}{Input}
\SetKwInput{KwInitialize}{Initialize}
\SetKwInput{KwOutput}{Output} 
\SetKwComment{Comment}{/* }{ */}
\usepackage{xcolor,soul,framed} 
\usepackage{xspace}
\colorlet{shadecolor}{yellow}
\usepackage[pdftex]{graphicx}
\graphicspath{{../pdf/}{../jpeg/}}
\DeclareGraphicsExtensions{.pdf,.jpeg,.png}

\usepackage{array}
\usepackage{url}
\usepackage{cite}

\usepackage{multirow}
\usepackage{multicol}
\usepackage{amsmath}
\usepackage{amssymb}
\begin{document}
    \title{Massive MIMO CSI Feedback using Channel Prediction: How to Avoid Machine Learning at UE?}
  \author{M. Karam~Shehzad,~\IEEEmembership{Member,~IEEE,}
      Luca~Rose,~\IEEEmembership{Member,~IEEE,}     and~Mohamad~Assaad,~\IEEEmembership{Senior Member,~IEEE}

  \thanks{M. Karam Shehzad is with Nokia Standards and CentraleSupelec, University of Paris-Saclay, Paris-Saclay, France. Luca Rose is with Nokia Standards, Massy, France. Mohamad Assaad is with CentraleSupelec, University of Paris-Saclay, Paris-Saclay, France. (e-mails: \{muhammad-karam.shehzad, luca.rose\}@nokia.com); mohamad.assaad@centralesupelec.fr.}
  \thanks{The corresponding author is M. Karam Shehzad.}
  \thanks{Channel prediction methods, proposed in this work, have been published at IEEE GLOBECOM 2022 \cite{karam_CP_Globecom}.}

}

\markboth{This article has been accepted for publication in IEEE Transactions on Wireless Communications
}{Muhammad \MakeLowercase{\textit{et al.}}: Channel Prediction for Massive MIMO CSI Feedback: How to Avoid Machine Learning at UE?}

\maketitle

\begin{abstract}
In the literature, machine learning (ML) has been implemented at the base station (BS) and user equipment (UE) to improve the precision of downlink channel state information (CSI). However, ML implementation at the UE can be infeasible for various reasons, such as UE power consumption. Motivated by this issue, we propose a CSI learning mechanism at BS, called CSILaBS, to avoid ML at UE. To this end, by exploiting channel predictor (CP) at BS, a light-weight predictor function (PF) is considered for feedback evaluation at the UE. CSILaBS reduces over-the-air feedback overhead, improves CSI quality, and lowers the computation cost of UE. Besides, in a multiuser environment, we propose various mechanisms to select the feedback by exploiting PF while aiming to improve CSI accuracy. We also address various ML-based CPs, such as NeuralProphet (NP), an ML-inspired statistical algorithm. Furthermore, inspired to use a statistical model and ML together, we propose a novel hybrid framework composed of a recurrent neural network and NP, which yields better prediction accuracy than individual models. The performance of CSILaBS is evaluated through an empirical dataset recorded at Nokia Bell-Labs. The outcomes show that ML elimination at UE can retain performance gains, for example, precoding quality.
\end{abstract}


\IEEEpeerreviewmaketitle

\section{Introduction}
\IEEEPARstart{A}{rtificial} Intelligence (AI) and Machine Learning (ML) have emerged as a paradigm shift for 5G-advanced (5G+) and 6G cellular networks \cite{Karam_AI}. AI and ML are promising techniques to solve non-convex optimization problems; hence, they can bring benefits to algorithm-deficit and model-deficit problems. For example, the amalgamate of ML and massive multiple-input multiple-output (MIMO) technology can enhance precoding gain \cite{CSI_TCCN_Karam}. In the literature, promising benefits of ML in a realistic environment have been observed, which show that it can pave the way for introducing intelligence in 6G networks \cite{AI_Real_World, Karam_CP}. However, in many instances, to attain the benefits of ML and massive MIMO, user equipment (UE) is required to run ML algorithm(s) \cite{Karam_AI, CS_FB_precoding_arxiv, Karam_Dealing, CSI_TCCN_Karam, DL_CSIFB, Karam_UAV}. Broadly, this paper proposes an alternative scheme for massive MIMO channel state information (CSI) feedback that circumvents the need for power-hungry ML implementation at the UE while maintaining the overall ML gains.

\subsection{CSI Feedback: Motivation and State-of-the-Art}
Massive MIMO is one of the promising technologies which can enhance the performance of 6G cellular networks \cite{TWC_Ref1}. The available research body on massive MIMO reveals the importance of such technology and its dependence on accurate and timely CSI acquisition \cite{CSI_FB_Rana, CSI_FB_Survey, CSI_FB_Reciprocity}. Specifically, CSI feedback overhead in a massive MIMO system is one of the fundamental issues yet to be addressed. ML-assisted CSI feedback can be a promising direction to reduce overhead. It is also proposed as one of the use cases in the 3rd Generation Partnership Project (3GPP) Release-18 physical layer -- AI/ML for New Radio Air Interface \cite{Karam_AI}.

The concept of reporting estimated CSI from UE to BS is termed as \textit{CSI feedback} \cite{CSI_FB_Nokia}. It is composed of three parts: Precoding Matrix Indicator (PMI), Channel Quality Indicator (CQI), and Rank Indicator (RI). PMI is the most important one, as it helps the BS to select an appropriate beam. Briefly, the transmitted CSI reference symbol (CSI-RS) that gives the best signal-to-interference-plus-noise ratio (SINR) is selected, then the corresponding PMI from codebook
\footnote{It is a complex-valued matrix that transforms the data bit into another dataset, mapping to each antenna port. In 5G, two types of codebook are defined: type-I and type-II \cite{3GPP_5G}. Type-I reports only the phase of the selected beam, whereas type-II reports subband and wideband amplitude information. In comparison, type-II is more detailed CSI reporting and is mainly designed for multiuser MIMO. It linearly combines a group of beams within a group, while type-I selects one beam from a group of beams.}
(a set of precoding matrices) is reported to the BS along with CQI
\footnote{Used for the indication of channel quality to BS. It has a value between $0$ to $15$, indicating the modulation and coding level that UE can operate.}
and RI\footnote{It is used to report the number of independent communication channels, which can help to understand how well multiple antennas work, i.e., is the signal transmitted by different antennas correlated or not?}.
With the goal of reducing over-the-air (OTA) overhead, estimated CSI is compressed, i.e., reporting PMI; thereby, acquired CSI at BS is prone to compression errors. Such compression can deteriorate the performance of a massive MIMO precoder. In this article, we will learn how implementing ML at BS can help reduce compression errors, OTA overhead, and the computation cost of UE.        

CSI feedback overhead increases with the number of antennas. Compression is one of the solutions used in the standards, but it degrades the performance of the estimated channel. In the literature, ML and deep learning (DL) have been applied to improve the precision of acquired CSI at BS. The authors of \cite{DL_CSIFB, CSI_FB_TWC1, CSI_FB_TWC2, CSIFB_new_1, CSIFB_new_2, CSI_FB_TWC3, CSIFB_new_3, CSI_FB_TWC4, CSIFB_new_5, CSI_FB_TWC5, Auto_Encoder_1, IEEE_WCL1, IEEE_WCL2, IEEE_WCL3, Auto_Encoder_2, Auto_Encoder_3, CS_FB_Arxiv, Auto_Encoder_4, Auto_Encoder_5, CSI_FB_TWC6} exploit \textit{autoencoder} for CSI feedback. Particularly, inherent features of \textit{autoencoder}, i.e., \textit{encoder} and \textit{decoder}, are used to compress and recover CSI, respectively. The former, implemented at UE, encodes the estimated CSI into codewords and the latter for reverse-engineering, i.e., retrieving the estimated CSI at BS from transmitted codewords. Nevertheless, the major disadvantages of \textit{autoencoder} are high training cost (e.g., huge dataset is required, massive number of parameter tuning, and model validation), imperfect decoding, misunderstanding of influencing variables, and preservation of irrelevant information \cite{Autoencoder_Disadvantages}. Lastly, collaboration between the BS and a UE is indispensable, posing new challenges for the standardization bodies \cite{CSI_FB_Survey}. A different approach has been proposed in \cite{Karam_KF, Karam_Dealing, CSI_TCCN_Karam}, where the CSI prediction technique has been exploited to enhance CSI quality. In a nutshell, twin channel predictors (CPs) are used at the BS and UE to enhance precoding gain and reduce OTA overhead. One of the limitations of the works proposed in \cite{Karam_Dealing} and \cite{Karam_KF} is the synchronization of twin CPs. Besides, reporting massive amounts of ML-based trained weights to the BS is another crucial issue, which causes high OTA overhead \cite{CSI_TCCN_Karam}. A common problem associated with state-of-the-art is the implementation of ML/DL at the UE. Thus, storage and training of massive neural network (NN) at a UE can be nearly infeasible.
\subsection{Our Contributions}
Motivated by the issues above, this paper considers a light-weight predictor function (PF) to reduce CSI feedback overhead and improve acquired CSI accuracy. More specifically, assuming the limited power of a UE, we address the question: \textit{How to remove ML from the UE while maintaining ML gains?} This question, to the best of our knowledge, has not been addressed in the literature.

In light of the above question, we consider the implementation of CSI prediction at the BS for massive MIMO CSI feedback, where a light-weight PF is computed by exploiting predicted and reported CSI realizations. Later, PF, reported by the BS, can be used at the UE for feedback evaluation. Hence, training and storage of a fully functional NN can be eliminated at the UE. PF reporting can help create a set of matrices, i.e., a codebook, which can have a standardization impact. Thus, an index can be reported to avoid the transmission overhead of PF. To this end, we also cover possible standardization points of the proposed idea\footnote{UE can also perform ML training for various reasons, e.g., BS does not have a dataset for training ML algorithm(s). This approach is subject to UE being capable of running ML \cite{CSI_TCCN_Karam}. Besides, CSILaBS can work without ML, assuming that the channel is evolving using an autoregressive (AR) process.}, called CSI learning at BS (CSILaBS).  

As CSI prediction is pivotal in CSILaBS, we also address ML-based CPs. Specifically, we exploit time-series models of DL, e.g., Bidirectional long short-term memory (BiLSTM). In addition, we use NeuralProphet (NP), a recently introduced time-series model composed of statistical components, e.g., AR, for CSI prediction. Furthermore, inspired by a statistical model, we develop a novel hybrid framework comprising a recurrent neural network (RNN), an ML algorithm, and an NP to achieve better prediction accuracy. In addition
to this, we employ hyperparameter tuning for each of these
individual models to select only the best training parameters. The performance of the proposed work is evaluated on a real-world dataset recorded at the Nokia Bell-Labs campus in Stuttgart, Germany. 

By extending the work to multiuser, we address feedback selection methodologies to acquire more accurate CSI at the BS by exploiting PF at the UE. Particularly, we propose different feedback selection methodologies while exploiting Glauber dynamics. We see the problem of CSI feedback selection as a random access scheme, but the goal of our study is different, i.e., we do not transmit at a given rate to improve throughput. We use random access schemes to reduce compression errors and enhance CSI feedback quality at the BS. Through simulations, we show that the proposed methodologies can effectively improve CSI precision when feedback is intelligently selected. We have learned from the extensive simulations that the prediction error threshold is an important feedback evaluation parameter in a multiuser environment that can be carefully selected to improve CSI.
\subsection{Paper Organization and Notations}
The rest of the paper is organized as follows. In Section\,\ref{system_model}, system model is presented. The proposed massive MIMO CPs are explained in Section\,\ref{channel_predictor}. CSI feedback scheme, CSILaBS, is detailed in Section\,\ref{CSILaBS}. The proposed feedback selection methodologies are addressed in Section\,\ref{FB_schedulers}. Section\,\ref{CSILaBS_Aspects} addresses the implantation aspects of CSILaBS. The description of utilized dataset is given in Section\,\ref{campaign}. Results are analyzed in Section\,\ref{Results}. Conclusion is made in Section\,\ref{Conclusions}.    
\subsubsection*{Notations}
Throughout this paper, matrices and vectors are represented by boldface upper and lower-case, respectively. Also, scalars are denoted by normal lower and upper-case.  
The superscripts $(\cdot)^{\dag}$, $(\cdot)^{*}$, and $(\cdot)^{\ddagger}$ denote the transpose, conjugate transpose, and pseudo-inverse of a matrix/vector, respectively. $\mathsf{E}\{\cdot\}$ denotes expectation operator, $\left\|\cdot\right\|^2_\text{FRO}$ represents squared Frobenius norm, and $|\cdot|$ shows absolute value. $\mathbf{h}_k$ denotes the true channel for $k^{th}$ UE, where $k=\{1,2, \cdots,K\}$. The acquired and predicted channel at the BS for $k^{th}$ UE are represented by $\bar{\mathbf{h}}^\text{BS}_k$ and $\widetilde{\mathbf{h}}^\text{BS}_k$, respectively. Similarly, the predicted and estimated channel at a UE are denoted by $\widetilde{\mathbf{h}}_k$ and $\widehat{\mathbf{h}}_k$, respectively. Furthermore, the notations $\mathbb{R}$ and $\mathbb{C}$ are representing the real and complex numbers, respectively.       
\section{System Model}\label{system_model}
We consider a massive MIMO cellular network, where a BS is serving using $M\gg1$ transmit antennas to $K$ single-antenna UEs. Without loss of generality, the received signal, per subcarrier and orthogonal frequency-division multiplexing (OFDM) symbol, at the $k^{th}$ UE can be expressed as
\begin{equation}
r_k(t)=\mathbf{h}_k(t)\boldsymbol{\nu}(t)+{\varpi}(t)\:,
\end{equation}
where $t$ is the time-index, $\mathbf{h}_k(t)\in\mathbb{C}^{M\times1}$ is the channel vector for $k^{th}$ UE,  $\boldsymbol{\nu}_k(t)=[\nu_1(t),\nu_2(t)\dots,\nu_{M}(t)]^\dag$ is the pilot vector, and ${\varpi}(t)$ is the noise. 

In the frequency-division-duplex system, $\mathbf{h}$ should be fed back to the BS by a UE, requiring high OTA overhead. However, $\mathbf{h}$ must be estimated at the UE before feedback \cite{DL_channel_estimation}. Channel estimation is beyond the scope of this study; therefore, we assume perfect channel estimation and focus on CSI feedback. In CSILaBS, we assume that a BS is equipped with an ML-based CP. Thus, in the following section, we first explain the proposed channel prediction methodologies and then address the proposed CSI feedback mechanism. Though ML-based channel prediction has already been considered in the literature, e.g., \cite{Karam_CP, CP_age} has many issues. We will address those issues in the following section, and later in Section\,\ref{Results}, we will show that the proposed channel prediction models outperform conventional methods.    
\section{Proposed Massive MIMO Channel Predictors}\label{channel_predictor}
Strictly speaking, channel prediction is the process of predicting future channel realizations by exploiting past observations \cite{CP_TWC1, CP_TWC2, CP_TWC3, CP_TWC4, CP_TWC5, CP_TWC6}. In the rest of this section, we highlight the channel prediction models\footnote{The proposed models are used to predict the channel of single-UE, and we leave extension to multiuser as future work.} that we used in our study \cite{karam_CP_Globecom}. They are divided into RNN, BiLSTM, and a hybrid model. In the following, we explain these models.
\subsection{Recurrent Neural Network}\label{RNN}
RNN has emerged as a promising technique for time-series predictions \cite{Karam_CP}. Due to recurrent components, its prediction capability surpasses feed-forward NNs. 
Focusing on wireless communications, RNN has been utilized for channel prediction in several recent works, e.g., \cite{Karam_CP}, where the prediction capability of RNN is verified using empirical data. 
In this study, we use RNN to design a hybrid model. Given the vastity of the available literature on RNN, \cite{Karam_CP}, we do not detail its structure. 
In summary, RNNs are fed with $d$-step delayed inputs and corresponding labels to generate multi-step ahead predicted CSI realizations. Later in Section\,\ref{Hybrid}, we will utilize the predicted channel vector of RNN to design the hybrid model.
\begin{figure*}[ht]
    \centering
    \subfigure[Single-cell of LSTM.]{\includegraphics[width=0.49\textwidth]{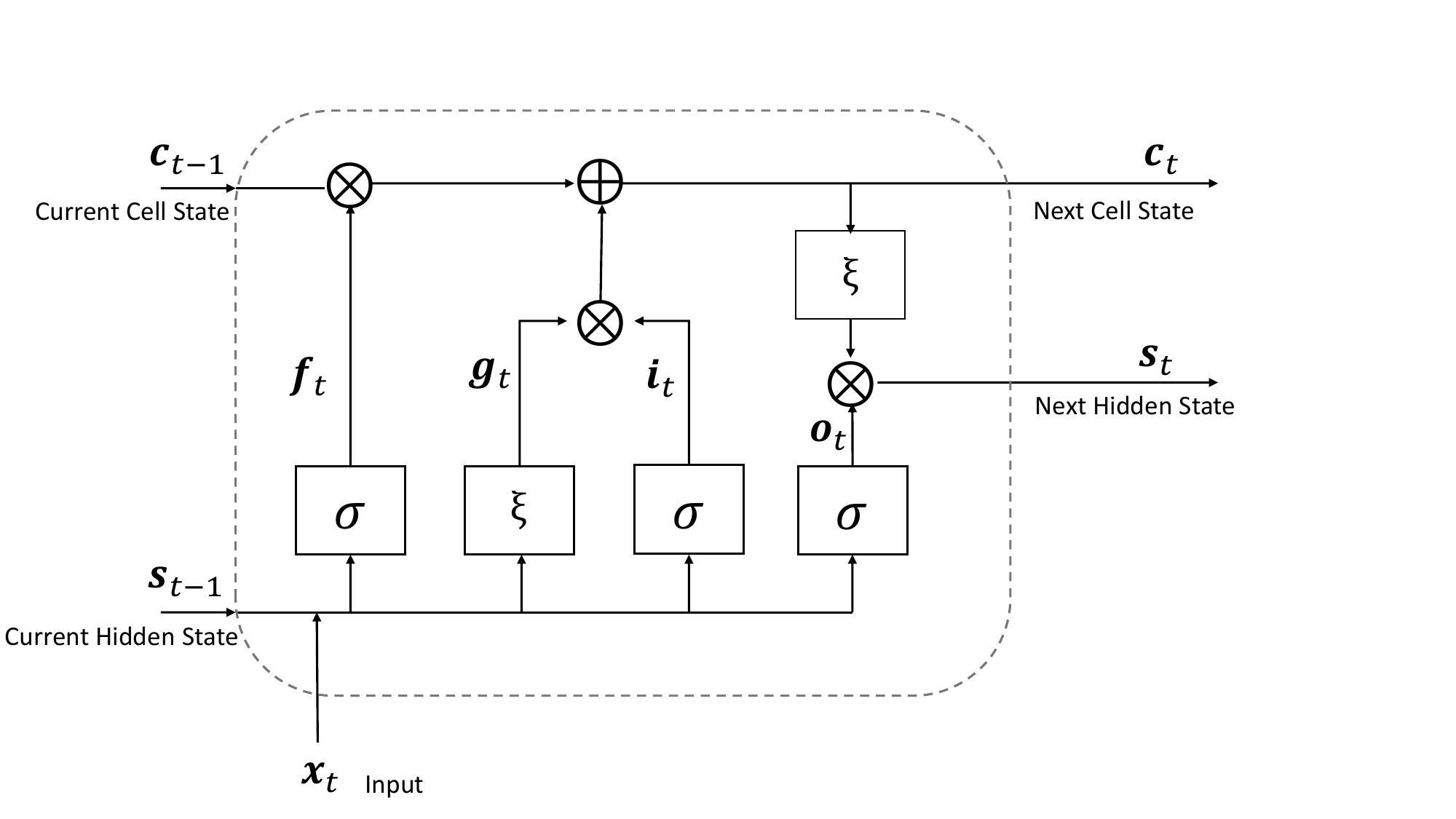}}
        \subfigure[Fully connected BiLSTM-based NN.]{\includegraphics[width=0.49\textwidth]{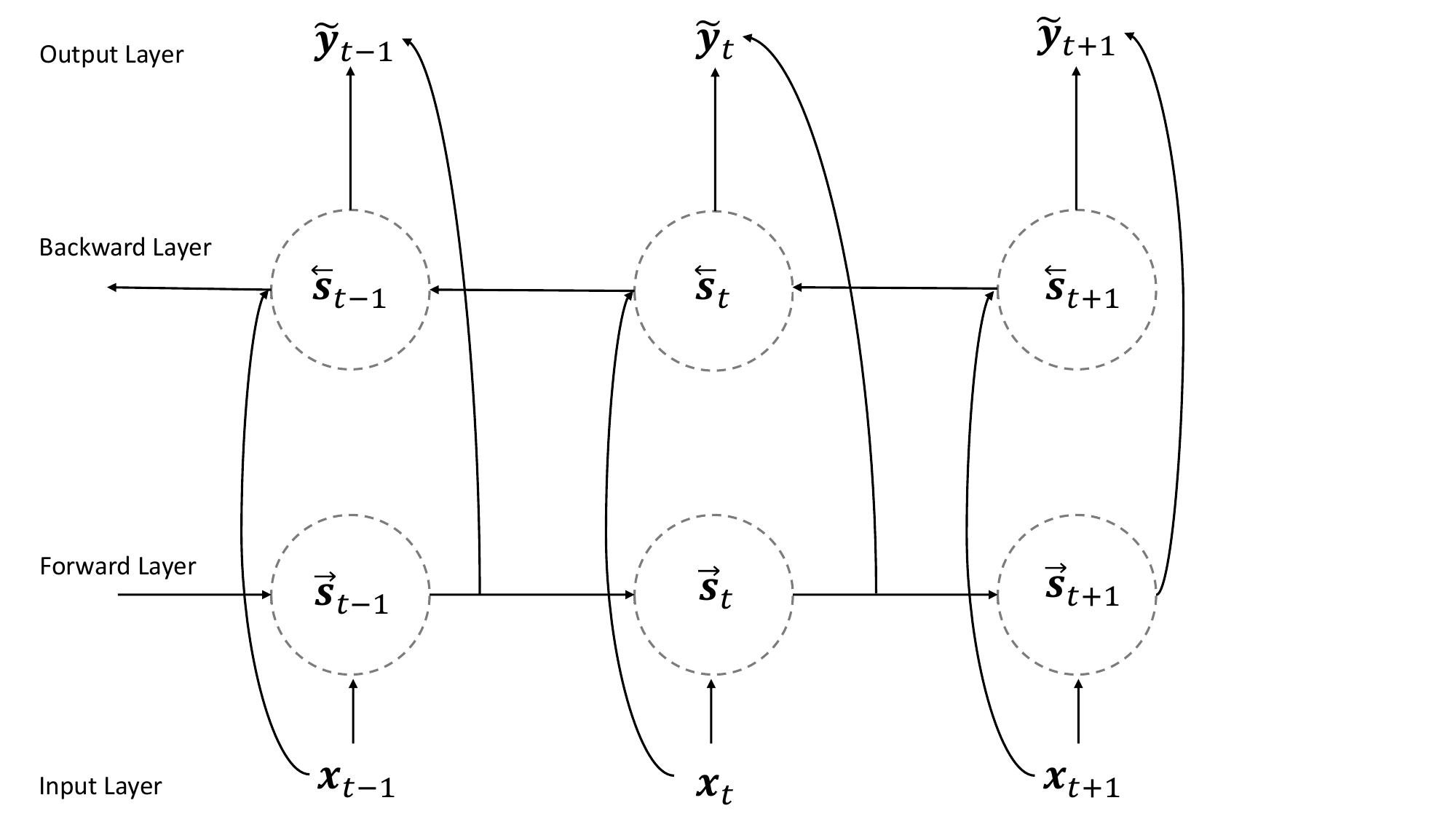}} \label{BiLSTM_a}
    \caption{Graphical illustration of time-series NNs, i.e., LSTM and BiLSTM. Fig.\,$1$\,(a) shows single-cell of an LSTM, where input is real-valued CSI realization for time instant $t$. Fig.\,$1$\,(b) depicts a fully connected BiLSTM-based NN, where inputs are learned in two ways. Circular shapes given in Fig.\,$1$\,(b) denote a single-cell of LSTM, which is given in Fig.\,$1$\,(a). The outputs of BiLSTM portray predicted CSI realizations. Source: \cite{karam_CP_Globecom}.}
    \label{LSTM-BiLSTM}
\end{figure*}
\subsection{Bidirectional Long Short-Term Memory}\label{BiLSTM}
BiLSTM is composed of two independent long short-term memory (LSTM) networks. In a BiLSTM model, information is learned from both ends of the input data vector, which results in better prediction performance than traditional unidirectional LSTM. LSTM is an advanced version of RNN. During model training, RNN suffers from vanishing and exploding gradient problems in the backpropagation. To solve this, an LSTM-based NN was developed. Schematic of an LSTM is depicted in Fig.\,\ref{LSTM-BiLSTM}. The core idea of LSTM is the introduction of a memory cell and multiplicative \textit{gates}, which regulate the flow of information. Briefly, \textit{forget gate} decides the amount of information to be stored in the cell by utilizing the current input, $\mathbf{x}_t$, and the output of the previous LSMT cell, denoted by $\mathbf{s}_{t-1}$. Mathematically,
    \begin{equation}
    \mathbf{f}_t= \sigma (\mathbf{W}_{f}\mathbf{x}_t+\mathbf{V}_{f}\mathbf{s}_{t-1}+\mathbf{b}_f)\:,\\
    \label{forget}
\end{equation}
where $\sigma(x)= \frac{1}{1+e^{-x}}$
is the \textit{sigmoid} activation function, $\mathbf{W}$ and $\mathbf{V}$ are the weight matrices, $\mathbf{b}$ is the bias vector, subscript $f$ is associated with the \textit{forget gate}.  
The \textit{input gate} determines the amount of information to be added into cell state $\mathbf{c}_{t-1}$ by exploiting $\mathbf{x}_t$ and $\mathbf{s}_{t-1}$. Mathematically,
     \begin{equation}
     \begin{aligned}
    \mathbf{i}_t &= \sigma (\mathbf{W}_{i}\mathbf{x}_t+\mathbf{V}_{i}\mathbf{s}_{t-1}+\mathbf{b}_i)\:,\\
        \mathbf{g}_t &= \xi (\mathbf{W}_{g}\mathbf{x}_t+\mathbf{V}_{g}\mathbf{s}_{t-1}+\mathbf{b}_g)\:,
        \label{input}
        \end{aligned}
\end{equation}   
where $\xi (\cdot)$ represents \textit{hyperbolic tangent} activation function, subscript $i$ and $g$ are associated with \textit{input gate}. By utilizing \textit{input} and \textit{forget gates},
LSTM can determine the amount of information to be retained and removed. Finally, the \textit{output gate} calculates the output of the LSTM cell by using an updated cell state, $\mathbf{c}_{t}$, and $\mathbf{x}_t$; the resultant output, given below, is then passed to the next LSTM cell of the network:  
    \begin{equation}
    \mathbf{o}_t= \sigma (\mathbf{W}_o\mathbf{x}_t+\mathbf{V}_{o}\mathbf{s}_{t-1}+\mathbf{b}_o)\\\:.
\end{equation}
As a result of the operations above, some information is dropped, and a few are added; this updates the next long-term state as follows:
\begin{equation}
    \mathbf{c}_t= (\mathbf{f_t}\otimes \mathbf{c}_{t-1}+\mathbf{i}_t\otimes \mathbf{g}_t)\:,
\end{equation}
where $\otimes$ represents Hadamard product. Lastly, short-term memory state, $\mathbf{s}_t$, is calculated by passing long-term memory, $\mathbf{c}_t$, through \textit{output gate} as
\begin{equation}
    \mathbf{s}_t= \mathbf{o}_t\otimes \xi (\mathbf{c}_t)\:.
\end{equation}

In BiLSTM architecture, as depicted in Fig.\,\ref{LSTM-BiLSTM}, input information is learned in two directions, i.e., left-to-right (forward layer) and right-to-left (backward layer). Importantly, notation $t+1$ in BiLSTM architecture is only used for illustration purposes; such indexes are based on passed CSI observations. The output information of each direction, denoted by $\vec{s}$ and $\cev{s}$, respectively, is passed simultaneously to the output layer, where predicted output is calculated as
\begin{equation}
    \widetilde{\mathbf{y}}_t=\vec{\mathbf{s}}_t\otimes\cev{\mathbf{s}}_t\:.
\end{equation}
\subsection{Hybrid Model}\label{Hybrid}
\begin{figure}[t]
\centering
\includegraphics[scale=0.27,trim=1cm 6cm 1cm 2cm]{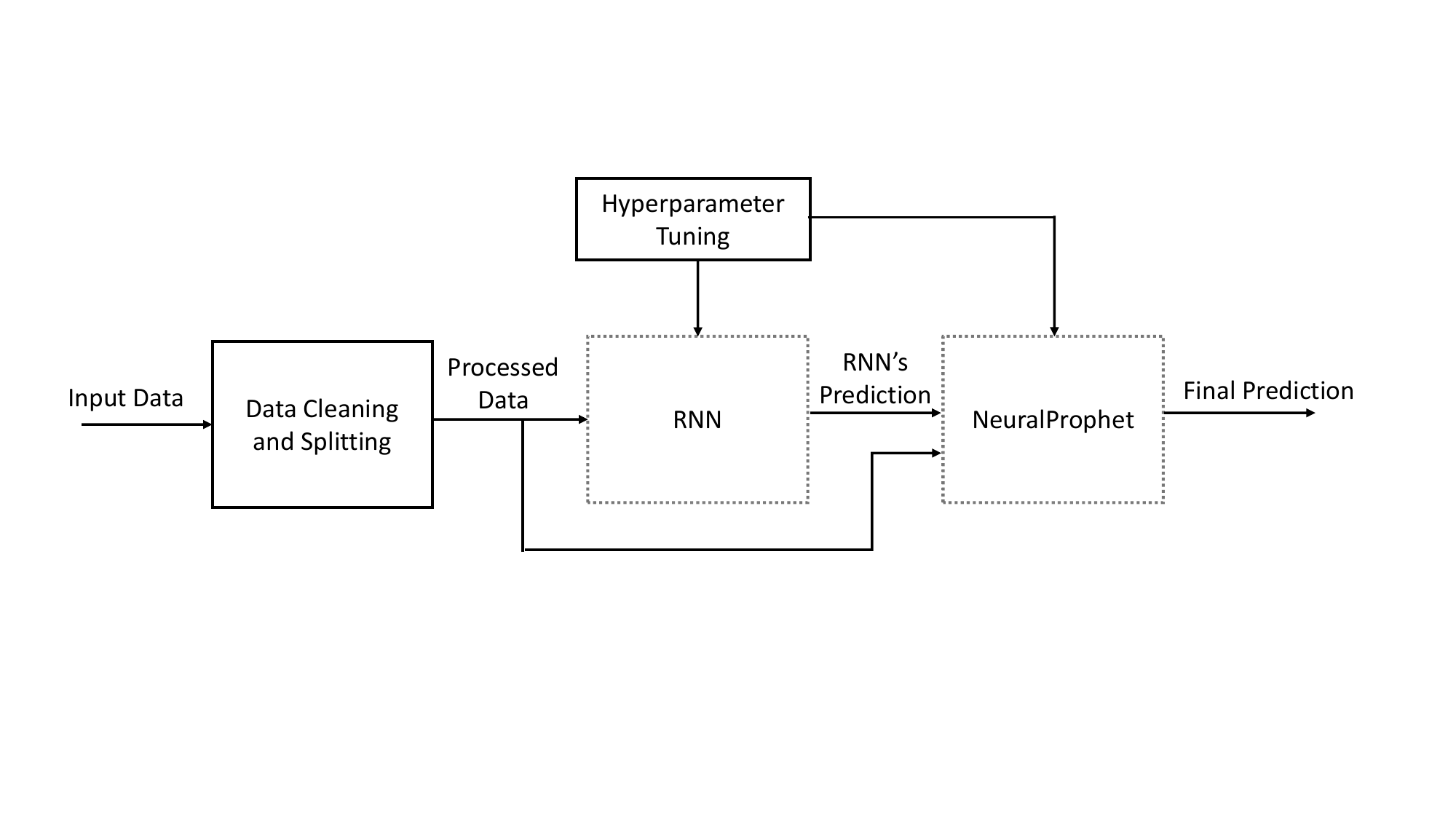}
\caption{Flow diagram of hybrid model, where dotted boxes are showing the prediction models used for the hybrid approach. Processed data is the real-valued CSI realizations, and final prediction denotes the multi-step ahead predicted CSI realizations. Source: \cite{karam_CP_Globecom}.}
\label{hybrid_model}
\end{figure}

In the hybrid model, as shown in Fig.\,\ref{hybrid_model}, we utilize an RNN-based CP, summarized in Section\,\ref{RNN}, and NP, explained in the following subsection. In the beginning, the dataset is cleaned, i.e., to check if there is corrupted/duplicate/missing data in the dataset. Additionally, data splitting is performed, in which the dataset is divided into three sets, i.e., training, validation, and testing. Further, the input sequences are transformed into an acceptable format, which ML models can process. The processed dataset is fed to the input of RNN and NP. Then, we consider the predicted channel vector of RNN\footnote{It is, however, important to mention that the predicted output of BiLSTM can also be considered. But for the sake of lower computationally complexity of the hybrid model, we use RNN.}, fed to NP along with input feature vector.
NP learns to correct the predicted output with RNN's prediction. Later in Section\,\ref{CP performance}, we demonstrate that NP can predict output more accurately when used with RNN and outperforms all standalone models. In the following, we explain the working functionality of NP. 
\subsubsection*{NeuralProphet}\label{NP}
Within a short span of time, NP has emerged as a promising technique for different time-series prediction tasks \cite{neuralprophet}. In the context of DL, several time-series models, e.g., RNN and BiLSTM, as explained above, have been developed. However, their internal functioning is still a question mark despite demonstrating promising results. In contrast, NP is an explainable, scalable prediction framework composed of statistical models, e.g., AR. It is sometimes important to analyze the performance of a prediction model in the form of different components. DL-based models are difficult to interpret due to their black-box nature. Contrarily, NP  is composed of different components\footnote{In the documentation of NP, it is composed of six components. However, with regard to our application of channel prediction, we dropped a few as some of them are irrelevant for CSI prediction, e.g., holidays. For more details, the interested reader can refer to \cite{neuralprophet}.}, where each component contributes additively to predicted output, and their behaviour is well interpretable. Further, each component is composed of individual inputs and modelling methodology. The output of each component is $D$-step ahead future CSI realizations. For the notational convenience, we will explain the model with $D=1$, which will be later extended for multi-step, i.e., $D$ future steps. In the context of channel prediction, the predicted value of NP for time instant $t$ can be written as \cite{neuralprophet}
\begin{equation}
    \widetilde{z}_t= R_t+ F_t+ A_t\:,
\end{equation}
where $R_t$ and $F_t$ represent trend and seasonality functions for input data, respectively. $A_t$ is the AR effect for time $t$ based on previous CSI realizations. Below, we explain each of them. 
\subsubsection*{$R_t$}The trend function captures the overall variation in the input data. It tries to learn the points where clear variation in the data occurs; these points are called change-points, represented by $\{n_1, n_2, \dots, {n_\mathsf{m}}\}$, composed of a total of $\mathsf{m}$ change-points (tuned using \textit{grid search} \cite{Grid_search}). A trend function can be expressed as
\begin{equation}
    R_t= (\zeta^0+(\mathbf{\Gamma}_t)^{\dag}\boldsymbol{\zeta})\cdot t+(\rho^0+(\mathbf{\Gamma}_t)^\dag\boldsymbol{\rho})\:,
\end{equation}
where $\boldsymbol{\zeta}=\{\zeta^1, \zeta^2, \dots, \zeta^{\mathsf{m}}\}$, and $\boldsymbol{\rho}=\{\rho^1, \rho^2, \dots, \rho^{\mathsf{m}}\}$, are the vectors of growth rate and offset adjustments, respectively, and $\boldsymbol{\zeta}\in\mathbb{R}^{\mathsf{m}\times 1}$ and $\boldsymbol{\rho}\in\mathbb{R}^{\mathsf{m}\times 1}$. Besides, $\zeta^0$ and $\rho^0$ are the initial growth rate and offset values, respectively. And, $\boldsymbol{\Gamma}_{t}=\{\Gamma^{1}_{t}, \Gamma^{2}_{t}, \dots, \Gamma^{\mathsf{m}}_{t}\}$, where $\boldsymbol{\Gamma}_t\in\mathbb{R}^{\mathsf{m}\times 1}$, which represents whether a time $t$ is past each change-point. For a $j^{th}$ change-point, $\Gamma^j_t$ is defined as
\begin{equation}
    \Gamma^j_t=
    \begin{cases}
      1, & \text{if}\ t\geq n_j \\
      0, & \text{otherwise}
    \end{cases}\:.
  \end{equation}
\subsubsection*{$F_t$} The seasonality function, modeled using Fourier terms, captures periodicity in the dataset and is expressed as
\begin{equation}
    F^{p}_t=  \sum_{o
=1}^{\digamma
} \bigg(a_o\cdot cos\bigg(\frac{2\pi ot}{p}\bigg)+b_o\cdot sin\bigg(\frac{2\pi ot}{p}\bigg)\bigg)\:,
\end{equation}
where $\digamma$ is the number of Fourier terms defined for seasonality with periodicity $p$. At a time step $t$, the effect of all seasonalities can be expressed as
\begin{equation}
    F_t= \sum_{p\in \mathbb{P}}F^p_t\:,
\end{equation}
where $\mathbb{P}$ is the set of periodicities. 
\subsubsection*{$A_t$} AR is the process of regressing a CSI future realization against its past realizations. The total number of past CSI realizations considered in the AR process is referred to as the order, denoted as $d$, of the AR process. A classic AR process of order $d$ can be modeled as
\begin{equation}
    a_t= q+ \sum_{e=1}^{d}\theta_e\cdot a_{t-e}+\epsilon_t \:,\label{AR_Model_1}
\end{equation}
where $q$ and $\epsilon_t$ are the intercept and white noise, respectively, and $\theta$ are the coefficients of AR process. The classic AR model can only make one-step ahead prediction, and to make multi-step ahead prediction, $D$ distinct AR models are required to fit. To this end, we use feed-forward NN and AR, termed as \textit{AR-Net} \cite{AR-Net}, to model AR process dynamics. NP-based \textit{AR-Net} can produce multi-step future CSI realizations by using one AR model. \textit{AR-Net} mimics a classic AR model, with the only difference of data fitting. \textit{AR-Net} is a feed-forward NN that maps the AR model. \textit{AR-Net} is designed so that the parameters of its first layer are equivalent to the AR-coefficients.   
 
In the \textit{AR-Net}, $d$ last observations of CSI realizations are given as input, denoted as $\mathbf{z}$, which are processed by the first layer and passed through each hidden layer. Correspondingly, $D$-step ahead future CSI realizations, denoted by $\widetilde{\mathbf{z}}=\{A_t, A_{t+1}, \dots, A_{t+D}\}$, can be obtained at the output layer. Mathematically, 

\begin{equation*}
\begin{aligned}
        \boldsymbol{\omega}^\text{out}_1&=\alpha(\mathbf{U}_1\mathbf{z}+\mathbf{b}^\text{NP}_1)\:,\\
        \boldsymbol{\omega}^\text{out}_\mathsf{i}&=\alpha(\mathbf{U}_\mathsf{i}\boldsymbol{\omega}^\text{out}_{\mathsf{i}-1}+\mathbf{b}^\text{NP}_\mathsf{i})\:,\quad\quad \text{for}\quad \mathsf{i}\in[2, 3, \dots, l]\\
        \mathbf{\widetilde{z}}&=\mathbf{U}_{l+1}\boldsymbol{\omega}^\text{out}_l\:,
\end{aligned}
\end{equation*}
where $\alpha(\cdot)$ is the \textit{rectified linear unit} (ReLu) activation function, written as
\begin{equation}
    \alpha(\gamma)=
    \begin{cases}
      \gamma, & \gamma\geq 0 \\
      0, & \gamma<0
    \end{cases}\:.
  \end{equation}
Further, $l$ is the number of hidden layers having $n_h$ hidden units in each layer, $\mathbf{b}^{\text{NP}}\in\mathbb{R}^{n_h\times 1}$ is the vector of biases, $\mathbf{U}\in\mathbb{R}^{n_h\times n_h}$ is the weight matrix for hidden layers, except for the first $\mathbf{U}_1\in\mathbb{R}^{n_h\times d}$ and last $\mathbf{U}_{l+1}\in\mathbb{R}^{D\times n_h}$ layers. 
In the AR component of NP, an important selection parameter is the order of AR, i.e., $d$, which is hard to select in practice. In general, $d$ is chosen such that $d=2D$, i.e., twice the number of the prediction horizon. For notational convenience, we express the predicted channel for $k^{th}$ UE by each model, i.e., RNN, BiLSTM, and hybrid, as $\widetilde{\mathbf{h}}_k(t+D)$, which we utilize for the design of CSILaBS in the following section.
\section{CSILaBS}\label{CSILaBS}
This section details the proposed CSI feedback mechanism, CSILaBS, by exploiting the predicted channel of the previous section. The core idea is to have the same PF at both ends, at BS and at each UE. Such PF will be generated at the BS and reported to all UEs along with CSI-RS. The UEs will compute an update function by exploiting PF and CSI-RS, which is then feedback to BS. Hence, a precise version of CSI can be acquired at the BS with minimum computation and model storage at the UEs. For the sake of simplicity, we explain CSILaBS for one UE, remarking that all UEs will follow the same process. CSILaBS provides an efficient and light-weight way for the UE and the BS to implement the same PF without having to report a massive number of ML weights, as followed in \cite{CSI_TCCN_Karam}. CSILaBS implementation involves six different stages: \textbullet\ ML training at BS \textbullet\ PF reporting \textbullet\ PF verification \textbullet\ CSI estimation \textbullet\ CSI compression and feedback \textbullet\ CSI retrieval. We explain them in the following. 
\subsection{ML Training at BS}
At the beginning of this stage, training data for ML-based CP is acquired at the BS. The training and prediction of CP have already been explained in Section\,\ref{channel_predictor}. For brevity, let us assume that the predicted channel at the BS for time instant $t$ is denoted by $\widetilde{\mathbf{h}}_k^\text{BS}(t)$, which we use to explain the rest of the scheme. 
\subsection{PF Reporting}\label{PF Report}
At this stage, BS reports a PF by exploiting the predicted channel from the previous stage. PF reporting can be of different forms, which we address below.   
\subsubsection{Model-Based Representation}
This method evaluates matrix-based models for channel evolution and returns the matrix composition. For example, an AR model of order $d$ can be adopted, already given in Equation\,\eqref{AR_Model_1}. For brevity, we rewrite in the form of reported PF as
\begin{equation}
{\widetilde{\mathbf{h}}}_k^\text{BS}(t)={\sum_{e=1}^{d}{\hat{\mathbf{F}}}_{k,e}(f_Q[\widehat{\mathbf{h}}_k}(t-e)])^\dag\:,
\end{equation}
where transition matrix $\hat{\mathbf{F}}$ (a PF) is reported to UE by the BS. $f_Q[\cdot]$ is a standard element-wise quantization/compression function, where the real and imaginary parts of the channel are separately compressed. The matrix (or set of matrices) can be reported using a codebook (similar to PMI reporting in type-I/II followed in the standards) and henceforth reporting only the index. Another way is to compress the matrix and therefore reporting $E_v({\hat{\mathbf{F}}})$, where $E_v(\cdot)$ could be, e.g., maximum eigenvector compression.
\subsubsection{Full Function Reporting}
In this methodology, an NN is converted to an equation. 
This is simply done by writing the input-output relationship, considering weights and activation functions. The methods to convert NNs into equations are known in the literature \cite{Karam_CP}. This is, however, a discouraged method, as the function itself can be too complex to be implemented and reported \cite{CSI_TCCN_Karam}. To reduce reporting overhead, the function index can be reported as follows. A certain number of functions are reported into a codebook, which will be set at a standard level. An NN is converted into an equation, and the closest function is selected in the set, i.e., giving the best prediction. In other words, this type of reporting is equivalent to considering ML at UE as proposed in \cite{CSI_TCCN_Karam}, with the only difference of ML training at BS. Therefore, in this study, we consider this function reporting as a benchmark scheme and call it MLaBE (ML at both ends).   
\subsection{PF Verification}\label{PF verify}
The BS may predict the CSI with sufficient accuracy; however, the generated PF is inaccurate due to dynamic network conditions. Thus, PF verification is essential before starting the CSI feedback mechanism. PF can be verified by reporting it to UE. For example, UE can compute the error based on reported PF and CSI-RS, using the former for CSI prediction and later for the latest CSI estimation. For instance, the error can be calculated as a mean-squared error between the predicted and estimated CSI. The outcome can be sent to BS, which can retrain the NN if the error is sufficiently larger than a threshold value. 
\subsection{CSI Estimation}\label{channel estimation}
If PF is nearly accurate, the BS can utilize the predicted channel for CSI acquisition. To this end, BS transmits CSI-RS at, for example, time instant $t$, where UE can estimate the channel by exploiting CSI-RS. We assume perfect channel estimation; in the following, we address CSI compression once the channel is estimated. 
\subsection{CSI Compression and Feedback}\label{CSI compression}
Let us assume that the estimated CSI at the $k^{th}$ UE is denoted by $\widehat{\mathbf{h}}_k(t)$. Further, by exploiting the reported PF, explained in Section\,\ref{PF Report}, predicted CSI at the $k^{th}$ UE is represented by $\widetilde{\mathbf{h}}_k(t)$. Finally, CSI can be feedback to BS in the following manner \cite{CSI_TCCN_Karam}
\begin{equation}
\bar{{\mathbf{h}}}_k(t)=f_Q\bigg[\Omega\{\widetilde{\mathbf{h}}_k(t),\widehat{\mathbf{h}}_k(t)\}\bigg]\:,
\label{compressed_update}
\end{equation}
where $\Omega\{\cdot\}$ is an update function, which is simply the difference between the predicted, $\widetilde{\mathbf{h}}_k$, and the estimated channel, $\widehat{\mathbf{h}}_k$. The benefit of such compression is reduced OTA overhead to feedback CSI. For example, if $\Omega\{\cdot\}=0$, which is subject to perfect prediction, then the feedback-related overhead is eliminated; hence, UE does not need to feedback anything. Alternatively, if $\Omega\{\cdot\}\ne0$ then quantization function, $f_Q[\cdot]$, will add less noise, thanks to $\widetilde{\mathbf{h}}_k(t)$. It is important to mention that the above equation can also serve as a verification for PF before compression because UE can check if the error is sufficiently large.
\subsection{CSI Retrieval}\label{CSI acquisition}
CSI for time instant $t$ can be acquired at the BS by exploiting predicted CSI at BS, $\widetilde{\mathbf{h}}_k^\text{BS}(t)$, and the reported update $\bar{{\mathbf{h}}}_k(t)$. If $\Omega\{\cdot\}\ne0$, then BS retrieves CSI of $k^{th}$ UE as \cite{Karam_KF}
     \begin{equation}
    \bar{\mathbf{h}}_k^\text{BS}(t)= \widetilde{\mathbf{h}}_k^\text{BS}(t)+\Omega^{-1}\{\bar{{\mathbf{h}}}_k(t)\}\:,
    \label{proposed equation}
\end{equation} 
where $\Omega^{-1}$ represents inverse of $\Omega$. In this study, $\Omega^{-1}=-\Omega$. In the second scenario, i.e., $\Omega\{\cdot\}=0$, BS receives nothing, indicating that the predicted CSI at BS is perfect. Thus, the BS assumes that $\bar{\mathbf{h}}_k^\text{BS}(t)=\widetilde{\mathbf{h}}_k^\text{BS}(t)$. 
\subsection*{Remarks on CSILaBS}
The advantage of CSILaBS is the elimination of ML at the UE\footnote{If training data is unavailable at BS or does not have sufficient resources due to certain reason, then UE can run ML at its end \cite{CSI_TCCN_Karam}, helping the BS not to do ML training. Furthermore, in the case of using ML at UE, UE will be responsible for reporting PF to BS for channel prediction at BS. In that case, PF verification, discussed in Section\,\ref{PF verify}, will be done locally at UE, and BS has only to make predictions by exploiting reported PF. This study aims to reduce the computation cost of UE; hence, ML is considered at BS. To reduce the reporting overhead of training data, data augmentation algorithm(s) can be considered at BS to generate training data \cite{data_augmentation}, and the verification of such data can be done by exploiting codebook, which we leave as future work.}. Similar to \cite{CSI_TCCN_Karam}, another advantage is the reduction in feedback overhead if the prediction is not good enough and the elimination of feedback overhead if the prediction is perfect. Considering that the accurate PF is achieved, CP at the BS can keep running in the background, using the newly acquired channel, $\bar{\mathbf{h}}^\text{BS}$, as an input. The CP will be updated accordingly or retrained if necessary. The BS will trigger an update in the PF whenever a threshold is surpassed, and the updated PF will be reported to UE, where codebook entries can be updated, etc. This allows to track sudden changes in the ML model. For instance, if the underlying channel evolution changes due to, e.g., UE passing behind a building.

The transmission bandwidth might cover multiple Physical Resource Blocks (PRBs) or subbands for high data rate UEs. In that case, PF has to be reported from BS to UE for each PRB/subband; hence, the related reporting overhead might get large on the Physical Downlink Control Channel (PDCCH). Generally, UEs might be scheduled per Transmission Time Interval (TTI) on different PRBs/subbands. In addition, channel variations might vary significantly depending on PRB. Therefore, a fast and flexible adaptation of the PRBs/subbands for which a PF is being reported is beneficial. A simple solution can then be two-step approach and include a bitmap for, e.g., $13$ subbands, which will define per TTI  those subbands (bitmap value equal to $1$); thus, a new PF will be reported in this time slot. That way, limiting the downlink overhead to the relevant PRBs is possible. Note that if the allocation does not change, we can define a further bit with the meaning ‘keep the previous bitmap’.

The last critical remark is related to the compression function, $f_Q[\cdot]$, at UE. Suppose $f_Q[\cdot]$ uses infinite overhead bits. In that case, compression under CSILaBS and without CSILaBS will give the same output. Hence, CSILaBS will not provide any advantage. Later in Section\,\ref{Results}, we will show that high overhead brings negligible gain in CSILaBS. As feedback information is always compressed to reduce overhead, CSILaBS is beneficial under low overhead. 
In the following, we propose different algorithms to select the feedback in the multiuser environment.
\section{Feedback Selection for CSILaBS}\label{FB_schedulers}
In the multiuser scenario of CSILaBS, the feedback selection from each UE becomes crucial. In fact, the BS cannot know beforehand what the UEs need to feedback. Therefore, the UEs must decide dynamically if they have to feedback or not. This must be done in a distributed way. Thus, the problem of CSI feedback selection is similar to a random access scheme, where an UE senses the \textit{channel} before data transmission. For example, multiple UEs can transmit simultaneously in a random access scheme. In the case when two neighbouring UEs transmit at the same time, then we say that their messages \textit{collide} with each other, resulting in degrading the SINR of the overall system. Therefore, the UEs require a distributed medium access control (MAC) for efficient selection such that their messages do not \textit{collide}. 

A plethora of research has been proposed on random access protocols, e.g., carrier sense multiple access protocols, which are an important class of MAC protocols because of their simplicity. They have been widely used in IEEE 802.11 wireless-fidelity. In random access protocols, one potential solution is to use Glauber dynamics. In Glauber dynamics, the UEs transmit with a given probability, e.g., $\frac{\exp(\widetilde{\omega}_l(t))}{1+\exp(\widetilde{\omega}_l(t))}$ as given in \cite{CSMA_1}, where $\widetilde{\omega}_l(t)$ is the weight of link $l$ at time slot $t$. In the literature, random access-based approaches have proven to be throughput optimal \cite{CSMA_1, CSMA_2}. 

In CSI feedback selection, one may also see the problem as a random access scheme, but the goal of our study is different, i.e., we do not transmit at a given rate to improve throughput. We aim to enhance the CSI feedback. Therefore, we focus on using a random access model to acquire precise CSI at the BS. In a nutshell, we use the random access model in CSILaBS to reduce the errors due to compression imposed at the UE and to minimize OTA overhead costs. In this section, we address the question: \textit{How to efficiently select the feedback while improving the CSI accuracy of the overall system?} To answer this, in the following, we propose various algorithms. 

\subsection{Probabilistic Feedback}
In this scheme, the feedback from each UE is evaluated with a certain probability, which depends on the error in the update function computed at the $k^{th}$ UE. Without loss of generality, let us denote the set of available resource blocks to feedback the CSI in the given system by $\mathcal{N}=\{1,2,\cdots,N\}$. In the probabilistic feedback, the $k^{th}$ UE selects $n\in \mathcal{N}$ resource block randomly and evaluates the feedback in the selected resource block with a probability
\begin{equation}\label{prob_formula}
\mathcal{P}^n_k=\frac{\exp(\Delta_k)}{1+\exp(\Delta_k)}\:,
\end{equation}
where $\Delta_k=\mathsf{E}\{\lvert{{{\widetilde{\mathbf{h}}_k(t)}}-{\widehat{\mathbf{h}}_k(t)}}\rvert\}$ is the error between the predicted and the estimated channel at the $k^{th}$ UE. From \eqref{prob_formula}, it can be observed that higher $\Delta_k$ will result in higher $\mathcal{P}^n_k$. This will help enhance CSI feedback as UEs with high errors can be given priority so that an accurate version of CSI can be retrieved at the BS. Importantly, feedback is evaluated only when $\Delta_k>\wp$, where $\wp$ is the error threshold. The constraint $\Delta_k>\wp$ is put due to the nature of \eqref{prob_formula}, e.g., a UE may select the feedback with $50\%$ success rate even when there is no need to feedback CSI; in other words, when $\Delta_k=0$. Later in Section\,\ref{Results}, we will show that the optimal value of $\wp$ is an important design parameter for feedback selection. 
\color{black}
\begin{figure}[t]
\centering
\includegraphics[scale=0.40]{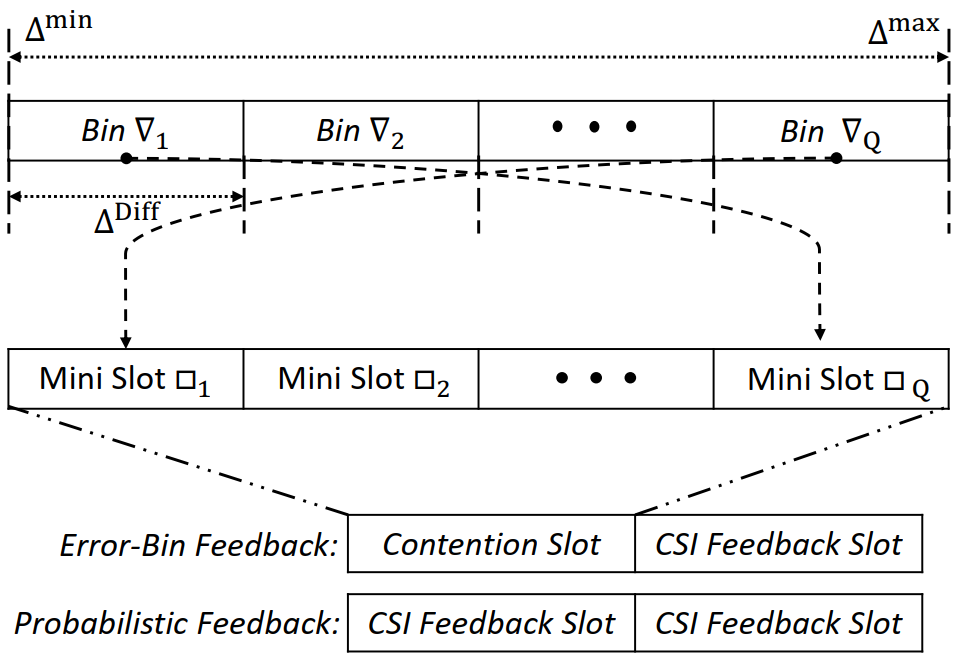}
\caption{Pictorial representation of two feedback resources in probabilistic and error-bin feedback. The former utilizes both resources for CSI feedback and later uses one to check contention and the second to feedback CSI of UE winning the contention.}
\label{Schedule_fig}
\end{figure}
\subsection{Error-Bin Feedback}
In the probabilistic feedback, we have learned that UEs with high error must be selected more frequently; hence, we revised the random access methodology accordingly. In the error-bin method, we tackle a similar problem as followed in the random access schemes, i.e., avoiding \textit{collisions}, but our objective is to acquire CSI at the BS with high precision. Considering the objective of acquiring precise CSI at the BS, we propose having a contention slot in addition to a data slot or CSI feedback slot. The contention slot can be further divided into mini-slots, where the objective of mini-slots is to prioritize CSI feedback of UEs having a high error. The UEs winning the contention will be selected to feedback CSI. To illustrate error-bin feedback, we have drawn Fig.\,\ref{Schedule_fig}, where the $k^{th}$ UE uses one resource to check contention and another to feedback CSI. The contention slot is composed of $\square_Q$ mini slots. Correspondingly, there are a total of $\nabla_Q$ bins, where each bin is equally spaced $\Delta^\text{Diff}$ and the error window of all bins is predefined, ranging between $\Delta^\text{min}$ and $\Delta^\text{max}$. For instance, $\Delta^\text{min}$ and $\Delta^\text{max}$ can be mapped between $0$ and $1$, respectively. And accordingly, $\Delta^\text{Diff}=0.10$.      

Focusing on the contention slot, the $k^{th}$ UE verifies the constraint $\Delta_k>\wp$ and feedback $1$ bit in a mini slot, depending on the error level. For example, if $\Delta_k$ lies in the first bin $\nabla_1$, i.e., $0<\Delta_k<0.10$, then the UE transmits $1$ bit in the last mini slot $\square_Q$ as the error is minimum. Similarly, if $\Delta_k$ belongs to the last bin $\nabla_Q$, showing high error, the UE feedback $1$ bit in the first mini slot $\square_1$. Fig.\,\ref{Schedule_fig} gives a pictorial illustration of these allocations. We can learn from this allocation that UEs with high errors will be given priority to feedback. Besides, when two or more UEs transmit in the same mini slot, this indicates \textit{collision}, but the advantage is we only lose $1$ bit. 

At the end of the contention slot, the UE winning will send the CSI in the CSI feedback slot. When two or more UEs win the contention, the UEs transmitted in the earliest mini slots will be selected. This is because the UE transmitting in the earliest slot indicates a high error. Notably, the selection of winning UEs depends on the number of available data slots. For instance, if there are two data slots and three UEs win the contention, the first two UEs with the highest error will be selected, and the third UE will be dropped due to the unavailability of the resource block. In this way, we prioritize UEs that have high errors. Consequently, CSI feedback performance can be enhanced. In contrast to error-bin feedback, probabilistic feedback uses all resources for CSI feedback, i.e., there is no contention slot (see Fig.\,\ref{Schedule_fig}).   

\subsection{Deterministic Feedback} 
In the deterministic feedback, all the UEs transmit at the same time, irrespective of their error value and without any resource selection. Intuitively, there will be many \textit{collisions}. Hence, CSI feedback performance cannot be enhanced as the BS will rely on open-loop CSI prediction, i.e. without any update from the UE. 
Later in Section\,\ref{Results}, we will show that the deterministic feedback performs the worst compared to error-bin and probabilistic feedback.     
\subsection{Periodic Feedback}\label{periodic}
This approach is used as a benchmark scheme to observe the performance gain of the proposed CSI feedback mechanisms. Briefly, we consider the feedback by exploiting the conventional CSI feedback scheme, i.e., without ML, where the feedback from a UE is simply the compressed version of the estimated channel. In other words, there is no channel predictor in the network; thus, $\bar{\mathbf{h}}_k^\text{BS}(t)=f_Q[\widehat{\mathbf{h}}_k(t)]$. The feedback without ML is evaluated in a round-robin fashion, i.e., every UE feedback CSI on its turn without any error threshold limit. For this purpose, the total UEs $K$ are divided into available resources. For instance, if there are $N=10$ resources and $K=30$, then $K=10$ UEs are selected in each round; hence, the first group of UEs will be able to retransmit in the fourth round. In Section\,\ref{Results}, we will discuss the results of the above CSI feedback selection mechanisms.
The pseudo-code of CSILaBS is given in Algorithm\,\ref{algo1}.
\begin{algorithm}[t!]
\label{algo1}
\DontPrintSemicolon
  \KwInput{$f_Q[\widehat{\mathbf{h}}_k]$}
  \KwOutput{{$\bar{\mathbf{h}}_k^\text{BS}(t) $}}
  {\tcp{\textcolor{black}{Step-1: Data aggregation at the BS. 
  }}}

{\tcp{Step-2: Training of CP at the BS by using one of the models given in Section\,\ref{channel_predictor}. Predict the channel ($\widetilde{\mathbf{h}}_k$) for time window $\mathsf{L}$.}}

{\tcp{\textcolor{black}{Step-3: Computation of PF ($\hat{\mathbf{F}}$).\;
Initialize variables:
$\mathbf{N}_\text{num}=\emptyset$
\:,$\mathbf{D}_\text{den}=\emptyset$}}}
\For{$\ell=1$ \textbf{to} $\frac{\mathsf{L}}{2}$}
  {
 \textcolor{black}{Select the entries of predicted channel, $\widetilde{\mathbf{h}}_k$, and corresponding true labels, $f_Q[\widehat{\mathbf{h}}_k]$.\;}
 $\mathbf{N}_\text{num}=[\mathbf{N}_\text{num}\,,\widetilde{\mathbf{h}}_k(\ell)]$\;
  $\mathbf{D}_\text{den}\:=[\mathbf{D}_\text{den}\,,f_Q[\widehat{\mathbf{h}}_k(\ell)]]$
  }
$\hat{\mathbf{F}}=\mathbf{N}_\text{num}\times(\mathbf{D}_\text{den})^{\ddagger}$\;
{\tcp{\textcolor{black}{Step-4: Verification and reporting of $\hat{\mathbf{F}}$ (see Section\,\ref{PF Report} and \ref{PF verify}).
}}}
{\tcp{\textcolor{black}{Step-5: Prediction at BS and UE using PF.\;
Initializing variables:
$\widetilde{\mathbf{H}}^\text{BS}_k=\emptyset$
\:,$\widetilde{\mathbf{H}}_k=\emptyset$
}}}
\For{$\ell=\frac{\mathsf{L}}{2}+1$ \textbf{to} $\mathsf{L}$}
  {
 \textcolor{black}{}
 $\widetilde{\mathbf{H}}^\text{BS}_k=[\widetilde{\mathbf{h}}^\text{BS}_k\,,\hat{\mathbf{F}}\times f_Q[\widehat{\mathbf{h}}_k(\ell)]]$\;
 \:$\widetilde{\mathbf{H}}_k\,=[\widetilde{\mathbf{h}}_k\,,\hat{\mathbf{F}}\times f_Q[\widehat{\mathbf{h}}_k(\ell)]]$
  }

{\tcp{\textcolor{black}{Step-6: Channel estimation at time $t$ by exploiting CSI-RS (see Section\,\ref{channel estimation}): $\widehat{\mathbf{h}}_k(t)$.}} }
{\tcp{\textcolor{black}{Step-7: The UE verifies locally if the acquired channel at the BS will be precise or not.}}}
{\tcp{\textcolor{black}{Step-8: Evaluate feedback decision using one of the algorithms used in Section\,V}}}
{\tcp{\textcolor{black}{Step-9: Channel acquisition at BS at time $t$ (see Section\,\ref{CSI compression} and \ref{CSI acquisition})}} }
      \Comment*[]{The following conditions holds true when the UE has feedback the CSI to the BS.}
  \eIf{$f_Q\{\Omega(\widetilde{\mathbf{h}}_k(t),\widehat{\mathbf{h}}_k(t))$\}==\text{true}}{
    $\bar{\mathbf{h}}^\text{BS}_k(t)= \widetilde{\mathbf{h}}^\text{BS}_k(t)+\Omega^{-1}\bigg[f_Q\{\Omega(\widetilde{\mathbf{h}}_k(t),\widehat{\mathbf{h}}_k(t))\}\bigg]$
  }
  {
     $\bar{\mathbf{h}}^\text{BS}_k(t)= \widetilde{\mathbf{h}}^\text{BS}_k(t)$
    }
\caption{\textcolor{black}{CSILaBS}}
\end{algorithm}
\section{Implementation Aspects}\label{CSILaBS_Aspects}
To implement CSILaBS, both the network entities, i.e., BS and UE, must assess the available capabilities. For example, the BS requests initial samples of the latest available CSI to initiate the proposed algorithm. Each UE has to report the training data to the BS, which may bring an additional overhead. Further, the BS has to verify the trained channel predictor by communicating with each UE. Furthermore, the BS has to report the light-weight PF to each UE, requiring overhead in the downlink. However, to reduce this overhead, we have discussed in Section\,IV that the codebooks can be established at the standard level; hence, an index from the codebook can be reported to predict the channel. In summary, a new CSI acquisition protocol comprising algorithm and memory requirements, length of the training dataset, and standardizing an ML algorithm for channel prediction, etc., is required to implement CSILaBS. 

The UEs typically do not stay in a cell, and the evolution of the underlying channel might change while moving to a different cell. In a highly dynamic environment, it is crucial to generalize an NN that can accurately predict the channel in different propagation conditions. Therefore, how to design an NN with high generalization is one of the crucial challenges in CSILaBS. One solution is to train the NN carefully with different underlying channel distributions. Another possible way is to perform online training; nevertheless, it must collect massive CSI samples, leading to an extra OTA overhead. 

As DL-based CSI feedback has been proposed as one of the use cases in 3GPP Release-18 AI/ML study items, the effect of CSILaBS on the existing standards needs to be evaluated. For instance, how much NMSE gain can be achieved through system-level and link-level simulations compared to type-I/II CSI feedback schemes? DL-based CSI feedback poses new challenges at the standardization level. To this end, the existing standards cannot be totally changed and can only be revised. For example, explicit feedback differs entirely from the current feedback framework and is complex to deploy in 5G-Advanced and 6G cellular networks. To further understand the implementation aspects of AI/ML on cellular networks, we refer to read \cite{Karam_AI}.  
\section{Dataset Description}\label{campaign}
\begin{figure*}[t]
\begin{center}
  \includegraphics[width=18cm, height=5cm]{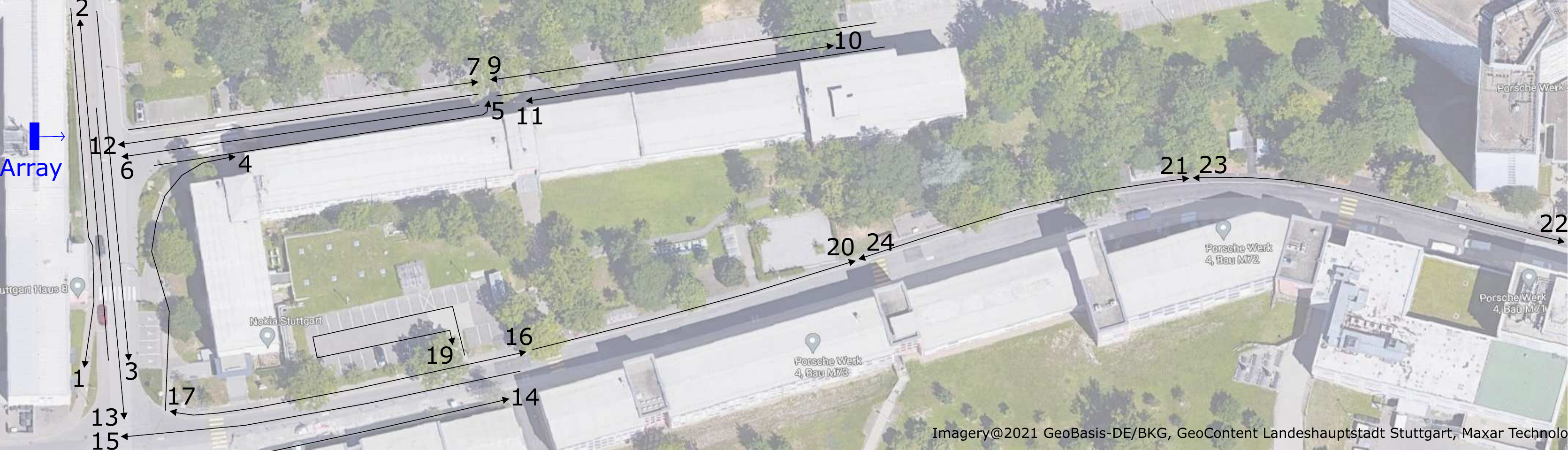}
\caption{{View of Nokia Bell-Labs campus in Stuttgart, Germany, where measurements are recorded. A mMIMO antenna, represented with a blue bar on extreme left-side of the figure, is located on a rooftop, transmitting in the direction represented with arrow. The UE is moved on the different tracks, illustrated with black lines, where arrowhead representing the direction of UE's movement and numerical values showing the track number. Source: \cite{Karam_CP}}.}\label{Map}      
\end{center}
\end{figure*}

\begin{figure*}
     \centering
     \subfigure[\textcolor{black}{Image of {Track-$1$ of the measurement campaign}.}]{\includegraphics[width=8cm,height=5.5cm]{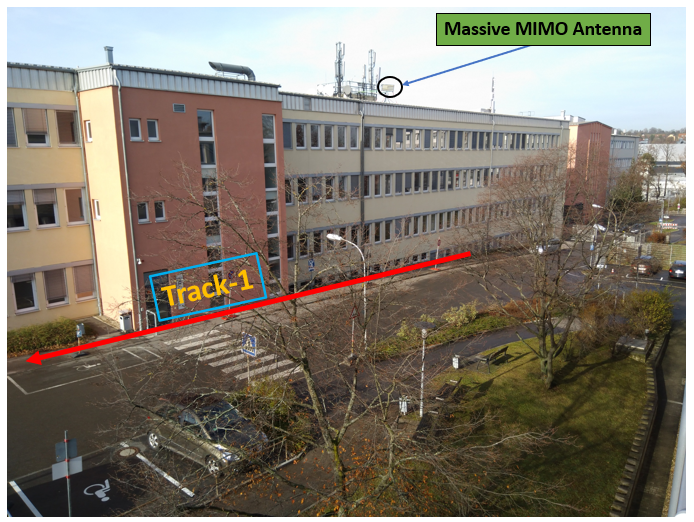}}{\label{route_1}}
     \subfigure[\textcolor{black}{BS antenna with $64$ antenna elements is shown here. }]{\includegraphics[width=6cm,height=5.5cm]{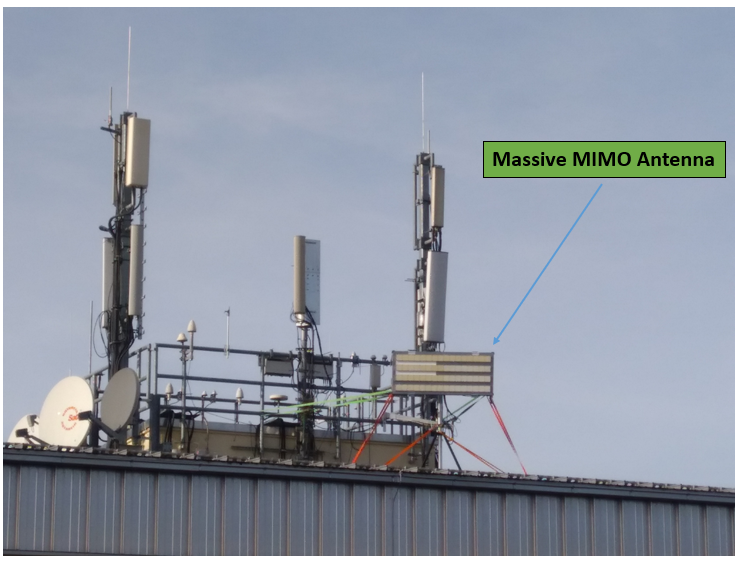}}{\label{antenna_1}}
     \subfigure[\textcolor{black}{A trolley, which is mimicking a UE, is moving on track-$1$.}]{\includegraphics[width=3cm,height=5.5cm]{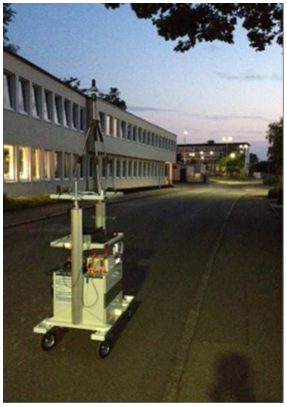}}{\label{UE_1}}
        \caption{\textcolor{black}{A closer view of measurement track-1 and network entities is shown in these figures. Buildings of approximately $15$\,m height can also be seen, which are acting as reflectors and blockers for the radio waves. Source: \cite{CSI_TCCN_Karam}}.}
        \label{track-1_BS_UE}
\end{figure*}

To observe the benefits of CSILaBS, a dataset is recorded in a practical wireless environment. To this end, a measurement campaign was performed at Nokia Bell-Labs campus in Stuttgart, Germany. The view of measurement campaign is shown in Figs.\,\ref{Map},\,\ref{track-1_BS_UE} \cite{Karam_CP, CSI_TCCN_Karam}. Briefly, a massive MIMO antenna, having $64$ antennas, is placed on a rooftop with approximately height of $15\,$m, and a UE equipped with a single monopole antenna having height $1.5\,$m is considered. The transmission is done using time-frequency orthogonal pilots at a carrier frequency of $2.18$\,GHz and using OFDM waveforms ($10\,$MHz long-term evolution (LTE) numerology, i.e., $600$ subcarriers with $15$\,kHz spacing). The UE is moved on different tracks, indicated in Fig.\,\ref{Map}, with a speed of $3$ to $5\,$km/h. The recorded data is composed of approximately 58\,seconds on each track, where $2000$ measurements are recorded at each second. More details of the dataset are available in \cite{Karam_CP}. In the case of multiuser, we exploit synthetic dataset, considering that the underlying model is evolving using an AR process.    
\section{Results and Analysis}\label{Results}
By exploiting the dataset given in Section\,\ref{campaign}, this section evaluates the performance of the proposed work. The results are divided into two parts, where first focuses on the performance of the proposed CPs and later evaluating the overall feedback scheme, CSILaBS. 

To train CPs, we used dataset of track-1 (shown in Fig.\,\ref{Map}), which is composed of $116$\,k consecutive CSI realizations, i.e., $\{\mathbf{h}(t)|t=1, \dots, 116\,\text{k}\}$. The dataset (normalized) is passed through necessary pre-processing and formatting steps using custom-built input pipelines so that it can easily parsed through each ML model. 
Further, the dataset of track-1 is divided into three parts: training (80\%), validation (10\%), and test (10\%). 
\begin{table} {}
\caption{Hyperparameters of ML Models}
\centering
 \begin{tabular}{|c| c| c| c|}
 \hline
 \textbf{Parameter} & \textbf{Value} & \textbf{Parameter} & \textbf{Value} \\ 
 \hline
 $\{l, \beta\}$ & $\{4, 1\}$ & $\{\gimel, n_h\}  \text{(NP)}$ & $\{0.001, 32\}$ \\ [1ex]
 \hline
 $\text{Epochs}$ & $50$ & $\{d,D\}$ & $\{20,10\}$ \\

 \hline
 $\mathsf{m}$ & $50$ & $\{\gimel, n_h\}\text{(RNN\:\&\:BiLSTM)}$ & $0.001, 200$  \\
 \hline
\end{tabular}

\label{table:I}
\end{table}
To train CPs, we use Huber loss as a cost function, which is defined as
\begin{equation}
    L_{\text{huber}}(\widehat{\mathbf{H}}_k,\widetilde{\mathbf{H}}_k)=
    \begin{cases}
      \frac{1}{2\beta}(\widehat{\mathbf{H}}_k-\widetilde{\mathbf{H}}_k)^2, & \text{for}\ |\widehat{\mathbf{H}}_k-\widetilde{\mathbf{H}}_k|\leq\beta \\
      |\widehat{\mathbf{H}}_k-\widetilde{\mathbf{H}}_k|-\frac{\beta}{2}, & \text{otherwise}
    \end{cases}\:
  \end{equation}
where $\widehat{\mathbf{H}}_k=\{\widehat{\mathbf{h}}_k({t+1}), \cdots, \widehat{\mathbf{h}}_k({t+D})\}$ are the true labels for the input CSI realizations $\widehat{\mathbf{X}}_k=\{\widehat{\mathbf{h}}_k({t-1}),\cdots,\widehat{\mathbf{h}}_k({t-d})\}$, and $\widetilde{\mathbf{H}}_k=\{\widetilde{\mathbf{h}}_k({t+1}), \cdots, \widetilde{\mathbf{h}}_k({t+D})\}$ is the $D$-step ahead predicted channel.  
By using Huber loss as a cost function, a batch of $32$ samples is fed into each CP, the predicted outcome is compared with true labels, and error is backpropagated to update weights and biases using \textit{adaptive moment estimation} (\textit{Adam}) as an optimizer.
The training iterations are repeated until cost function goes below threshold value.  
Furthermore, open-source libraries such as \textit{TensorFlow}, \textit{Keras}, and \textit{Scikit-learn}, are used for implementation of CPs. 

Selection of optimal training parameters, e.g., learning rate $(\gimel)$, number of hidden layers $(l)$, and hidden neurons $(n_h)$, play a pivotal role to enhance prediction accuracy. As hyperparameters are not directly learned by a model, rather, we manually define them before fitting the model. Therefore, a well-known automated strategy, that is, \textit{grid search} \cite{Grid_search}, is used for hyperparameter tuning.  
For NP, \textit{linear growth} function and $95\%$ \textit{change-point range} gave best results. Also, seasonality function is set to \textit{false} opposite to what is followed in \cite{karam_CP_Globecom}, remaining tuned parameters for NP and other CPs are listed in Table\,\ref{table:I}. Also, the standard NP model only takes uni-variate data to produce its output. In our hybrid model, we adapted the standard NP model for multivariate data by adding an extra future regressor into the NP model. Future regressors are the variables which are known for the future. In our case, predictions of RNN are known to us, and we included this information as a future regressor of the NP model. Thus, we feed both RNN predictions and true sequences as multivariate inputs to the NP model to enhance its prediction capability.
Besides, in RNN and BiLSTM models, the dropout layer is adapted, which drops hidden units randomly with a probability of $0.2$, to prevent over-fitting. 
\subsection{Performance of Channel Predictors}\label{CP performance}
The performance of four prediction models, i.e., NP, RNN, BiLSTM, and hybrid model, is evaluated by using normalized mean-squared error (NMSE) and cosine similarity. 
NMSE of a CP, for a single track, is defined as
\begin{equation}
\Upsilon^\texttt{CP}=\mathsf{E}\left\{\frac{\left\|{\widehat{\mathbf{H}}_k-\widetilde{\mathbf{H}}_k}\right\|^2_\text{FRO}}{\left\|\widehat{\mathbf{H}}_k\right\|^2_\text{FRO}}\right\}\:.
    \label{NMSE_Formula_CP}
\end{equation}
Similarly, by exploiting ${\widehat{\mathbf{H}}_k}$ and ${\widetilde{\mathbf{H}}_k}$, cosine similarity of a CP, denoted by $\varrho^\texttt{CP}$, is computed.  
\begin{figure}
\centering
\includegraphics[scale=0.39]{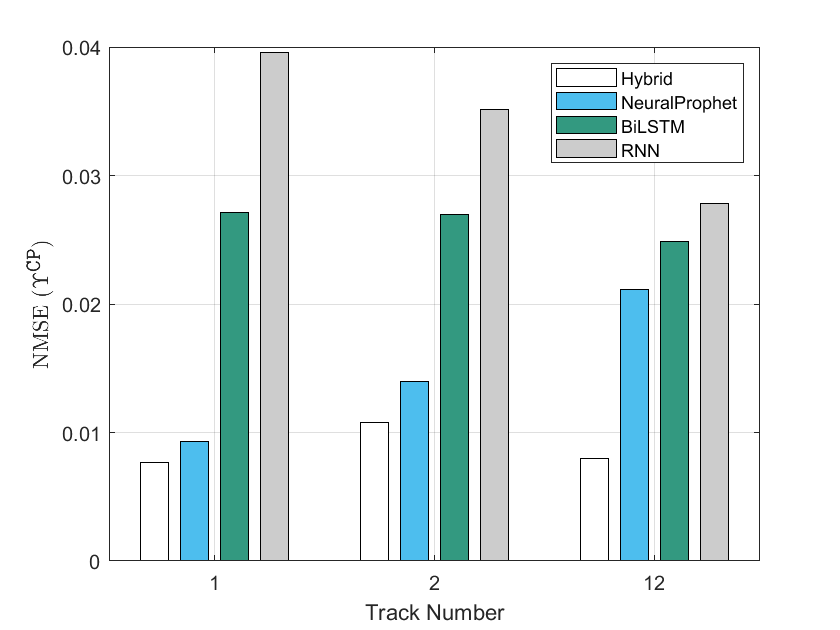}
\caption{Performance of the CPs on different tracks followed by UE, where NMSE is independent apropos track number as each track has different channel strength. The number of predicted CSI realizations is $D=10$, and uncompressed CSI is used for prediction. }
\label{NMSE_G1}
\end{figure}

\begin{table*}[t]
\begin{center}
\caption{Performance comparison of different channel predictors under various parameters}
\begin{tabular}{| *{10}{c|} } 
\hline

\multirow{2}{*}{\textbf{Overhead Bits\,$\downarrow$}} &\textbf{Prediction Model}\,$\rightarrow$& \multicolumn{2}{c|}{\textbf{Hybrid}}
& \multicolumn{2}{c|}{\textbf{NP}}& \multicolumn{2}{c|}{\textbf{BiLSTM}}
& \multicolumn{2}{c|}{\textbf{RNN}}\\
    \cline{2-10}

&\textbf{Prediction Horizon}\,$\rightarrow$&   ${D=10}$  &   ${D=1}$  &   ${D=10}$  &   ${D=1}$  &   ${D=10}$  &   ${D=1}$  &   ${D=10}$  &   ${D=1}$  \\

\hline
\multirow{2}{*}{$\infty$} & $\Upsilon^\texttt{CP}$& $\mathbf{0.0077}$&${0.0022}$&$0.0093$& $\mathbf{0.0014}$&$0.0271$&$0.0052$&$0.0396$&$0.0195$\\  
\cline{2-10}
&{$\varrho^\texttt{CP}$}&	$\mathbf{0.9975}$&$0.9993$&$0.9974$&$\mathbf{0.9996}$&$0.9923$&$0.9988$&$0.9857$&$0.9934$\\  
\hline

\multirow{2}{*}{$5$} & $\Upsilon^\texttt{CP}$& $\mathbf{0.0337}$&$0.0081$&$0.0413$& $\mathbf{0.0080}$&$0.0398$&$0.0111$&$0.0458$&$0.0254$\\  
\cline{2-10}
&$\varrho^\texttt{CP}$&	$\mathbf{0.9895}$&$0.9971$&$0.9864$&$\mathbf{0.9972}$&$0.9874$&$0.9964$&$0.9845$&$0.9914$\\  
\hline

\multirow{2}{*}{$3$} & $\Upsilon^\texttt{CP}$& $0.2188$&$0.0267$&$0.1550$& $\mathbf{0.0259}$&$\mathbf{0.0502}$&$0.0281$&$0.0586$&$0.0710$\\  
\cline{2-10}
&$\varrho^\texttt{CP}$&	$0.9416$&$0.9904$&$0.9435$&$\mathbf{0.9907}$&$\mathbf{0.9838}$&$0.9902$&$0.9808$&$0.9751$\\  
\hline

\multirow{2}{*}{$1$} & $\Upsilon^\texttt{CP}$& $0.2844$&$0.0242$&$0.1483$& $\mathbf{0.0241}$&$\mathbf{0.0278}$&$0.0303$&$0.0825$&$0.0263$\\  
\cline{2-10}
&$\varrho^\texttt{CP}$&	$0.9154$&$0.9899$&$0.9379$&$\mathbf{0.9899}$&$\mathbf{0.9886}$&$0.9876$&$0.9673$&$0.9890$\\  
\hline
\end{tabular}

\end{center}
\label{table:III}
\end{table*}

Fig.\,\ref{NMSE_G1} shows the comparison of four CPs in terms of NMSE $(\Upsilon^\texttt{CP})$. Particularly, performance is evaluated on three different tracks, which were unseen by the CPs during the training phase. The trend reveals that RNN, when used standalone, does not perform well; hence, it gives the worst performance. On the other hand, BiLSTM outperforms RNN, which is due to the fact that BiLSTM can retain information in their memory for longer time periods and learns the input data in both directions, thereby performing better. In contrast, the hybrid model performs better than other CPs. The rationale behind this is that NP learns better when RNN's predicted values are passed as future regressor. For instance, on track-1, there is approximately $80\%$ reduction in NMSE when hybrid model is used in comparison to RNN.

Table\,\ref{table:III} presents a detailed comparison of all CPs, where best results are written in bold numbers. The results are obtained for different prediction horizons, i.e., $D$, and under different quantization bits or overhead bits used to compress CSI. NMSE $(\Upsilon^\texttt{CP})$ and cosine similarity $(\varrho^\texttt{CP})$ are the two evaluation parameters used to analyze the performance of CPs. The results show that NP is the best among all for small prediction horizon, i.e., $D=1$, and under any compression level. However, for long prediction horizon, i.e., $D=10$, hybrid model is useful when compression level is low, e.g., $5$ or no-compression ($\infty$). The rationale behind the superior performance of hybrid model is the combination of RNN and NP. For instance, in case of $D=10$, NP has shown bad performance when used standalone but when combined with RNN, then it learned the information of RNN's prediction to improve accuracy. That is why hybrid model has superior performance. Nonetheless, under high compression and $D=10$, BiLSTM is the best choice, which is due to the fact that BiLSTM is suitable for long prediction horizons and handling non-linearities, thanks to memory cell and multiplicative \textit{gates} (see Section\,\ref{BiLSTM}).                 
\subsection{Performance of CSILaBS}
In this subsection, we compare the performance of CSILaBS, with two other schemes, i.e., MLaBE and without ML. In the case of without ML, $\bar{\mathbf{h}}^\text{BS}_k(t)=f_Q[{\widehat{\mathbf{h}}_k(t)}]$, and in CSILaBS and MLaBE, $\bar{\mathbf{h}}^\text{BS}_k(t)=\widetilde{\mathbf{h}}^\text{BS}_k(t)+\Omega^{-1}\{\bar{{\mathbf{h}}}_k(t)\}$. For brevity, we consider RNN as a CP for CSILaBS and MLaBE. For an abuse of notation, we denote $\bar{\mathbf{H}}^\text{BS}\in\mathbb{C}^{M\times K}$ and ${\mathbf{H}}\in\mathbb{C}^{M\times K}$ as the matrix form of $\bar{\mathbf{h}}_k^\text{BS}$ and ${\mathbf{h}}_k$, respectively, for all $K$ UEs, and $D=1$. By using this information, precoding gain can be obtained as
\begin{equation}
    	\Psi^\texttt{FB}= \mathsf{E}\left\{\text{Trace}(\mathbf{H}_\text{eq}^{*}\times \mathbf{H}_\text{eq})\right\}\:,
\end{equation}
where 
\begin{equation}  \mathbf{H}_\text{eq}=\left(\frac{\bar{\mathbf{H}}^\text{BS}}{\left\|{{{{\bar{\mathbf{H}}^\text{BS}}}}}\right\|_\text{FRO}}\right)^{*}\times \left(\frac{\mathbf{H}}{\left\|{{\mathbf{H}}}\right\|_\text{FRO}} \right)
\end{equation}
is the equivalent channel. Similarly, by using $\bar{\mathbf{H}}^\text{BS}$ and $\mathbf{H}$, NMSE, denoted by $\Upsilon^\texttt{FB}$, is calculated as
\begin{equation}
\Upsilon^\texttt{FB}=\mathsf{E}\left\{\frac{\left\|{\bar{\mathbf{H}}^\text{BS}-\mathbf{H}}\right\|^2_\text{FRO}}{\left\|\mathbf{H}\right\|^2_\text{FRO}}\right\}\:.
    \label{NMSE_Formula_FB}
\end{equation}
Similarly, cosine similarity, denoted by $\varrho^\texttt{FB}$, is calculated by using $\bar{\mathbf{H}}^\text{BS}$ and $\mathbf{H}$.  
\begin{figure}
\centering
\includegraphics[scale=0.39]{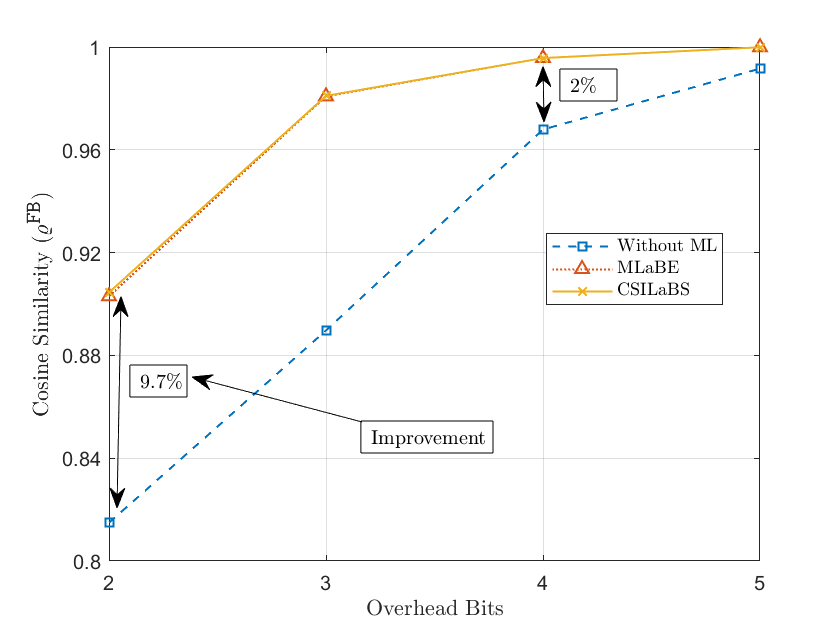}
\caption{Cosine similarity ($\varrho^\texttt{FB}$) of the acquired channel at BS and corresponding true channel. Figure demonstrates that $\varrho^\texttt{FB}$ increases with the number of overhead bits used to feedback channel from UE to BS.}
\label{CS_T1}
\end{figure}
\begin{figure}
\centering
\includegraphics[scale=0.39]{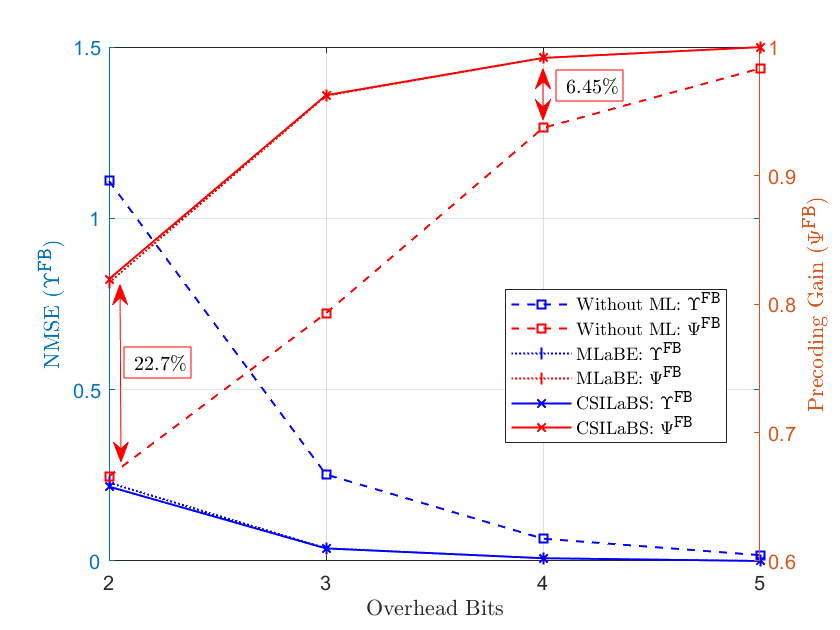}
\caption{NMSE ($\Upsilon^\texttt{FB}$) and precoding gain ($\Psi^\texttt{FB}$) of different CSI feedback schemes. Similar to Fig.\,\ref{CS_T1}, this figure reveals that performance of feedback CSI improves with the number of overhead bits. In other words, precise CSI acquisition comes at a cost of high overhead.}
\label{PG_NMSE_T1}
\end{figure}

Fig.\,\ref{CS_T1} presents the accuracy of the acquired channel at BS in the form of cosine similarity $(\varrho^\texttt{FB})$. The trend shows that the accuracy of the acquired channel increases with the number of overhead bits used to feedback CSI. However, CSILaBS brings a huge performance gain when feedback information is highly compressed, i.e., when $2$ bits are used. Numerically, CSILaBS shows approximately $11\%$ improvement in comparison to without ML. Thus, CSILaBS is beneficial in practical wireless communications environments as feedback evaluation methods, i.e., type-I/II, followed in the standards, are massively compressed due to exploitation of the codebook. Further, CSILaBS retains ML gain, i.e., CSILaBS gives nearly same gain as using ML at both ends (MLaBE). Nevertheless, the benefit of CSILaBS is the elimination of ML training at UE, and no need to store NN, which can be costly in terms of UE's power consumption. Also, CSILaBS uses a light-weight PF at UE for feedback evaluation, and overhead cost of PF reporting is small in comparison to MLaBE. For example, MLaBE requires $0.35\,$M parameters to report, whereas CSILaBS requires only $M\times M=4096$.      

Recalling that feedback CSI is used for precoding, we now inspect the performance of CSILaBS by using precoding gain, $\Psi^\texttt{FB}$, plotted on right-side of Fig.\,\ref{PG_NMSE_T1}. Furthermore, comparison is done by using NMSE, $\Upsilon^\texttt{FB}$, as evaluation parameter. Fig.\,\ref{PG_NMSE_T1} reveals that the CSILaBS gives massive precoding gain under high compression scenario. For example, in case of $2$ overhead bits, CSILaBS gives a gain about $22.7\%$ than without ML. This gain reduces with the number of overhead bits, e.g., $4$ overhead bits give a gain of only $6.45\%$. On the other hand, a similar trend can be observed for $\Upsilon^\texttt{FB}$. Further, MLaBE has comparable performance with CSILaBS, nevertheless, former requires implementation of ML at UE too.
\begin{figure}
\centering
\includegraphics[scale=0.39]{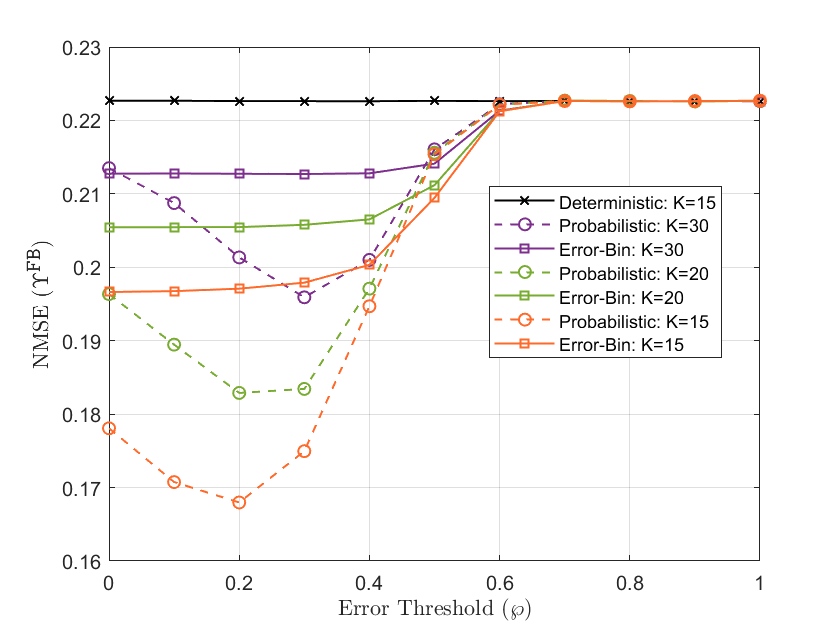}
\caption{NMSE $(\Upsilon^\texttt{FB})$  of different feedback schemes versus the error threshold $(\wp)$. A comparison is provided with various number of UEs $(K)$. $N=10$.}
\label{NMSE_FB_Ks_S10}
\end{figure}

Fig.\,\ref{NMSE_FB_Ks_S10} reveals the performance of $\Upsilon^\texttt{FB}$ against the error threshold $(\wp)$ selected to evaluate the feedback at a UE. It is important to mention that lower values of $\wp$ result in high numbers of feedback and vice versa. A comparison is provided with various numbers of UEs $(K)$ and different selection algorithms. First, we can observe that $\Upsilon^\texttt{FB}$ by using deterministic method gives the worst performance, which is due to excessive number of collisions. Hence, the BS relies on the predicted channel, i.e., $\bar{\mathbf{H}}^\text{BS}=\widetilde{\mathbf{H}}^\text{BS}$. Also, $\Upsilon^\texttt{FB}$ remains unchanged over different values of $\wp$, which is because $\bar{\mathbf{H}}^\text{BS}=\widetilde{\mathbf{H}}^\text{BS}$ is independent of $\wp$. In contrast, error-bin has slightly lower $\Upsilon^\texttt{FB}$, which increases with $\wp$ as feedback frequency has reduced when $\wp$ is increased. Contrarily to error-bin and deterministic feedback, probabilistic feedback is showing the best performance. This is due to the fact that the probabilistic feedback selects the feedback intelligently, i.e., when $\Delta$ and $\mathcal{P}$ are high. It can also be observed that with appropriate value of $\wp$, CSI acquisition can be improved. For example, when $K=30$ then $\wp=0.30$ is the optimal choice to select the feedback using probabilistic method. Numerically, we observe a gain of approximately $8.7\%$ and $13\%$ in comparison to error-bin and deterministic feedback, respectively. And this gain increases with the reduction in number of UEs as lower value of K results in less collisions. Importantly, in the case of deterministic feedback, $\Upsilon^\texttt{FB}$ will remain constant despite different values of $K$; hence, we only plotted for $K=15$ for the illustration purpose.                      
\begin{figure}
\centering
\includegraphics[scale=0.39]{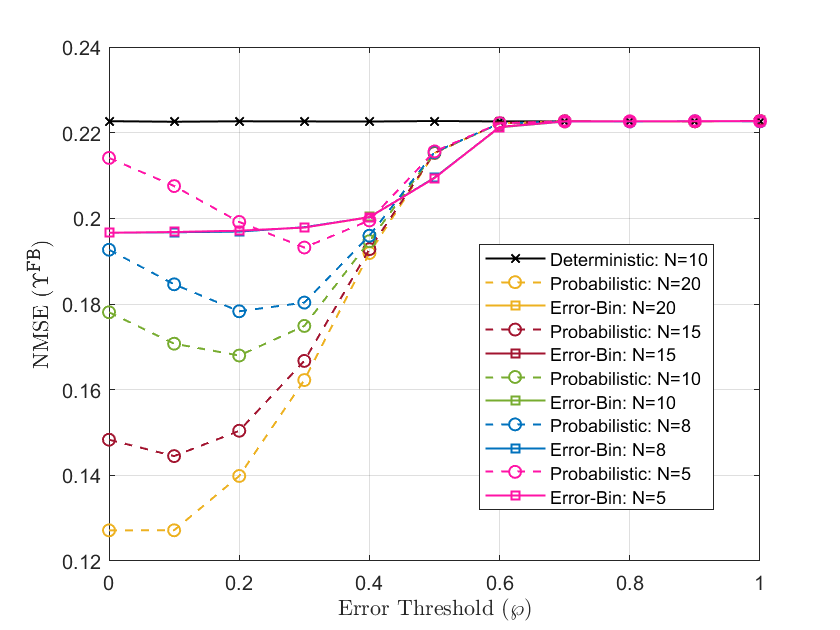}
\caption{Comparison with various number of resource blocks $(N)$. $K=15$.}
\label{NMSE_FB_Ss_K15}
\end{figure}

Fig.\,\ref{NMSE_FB_Ss_K15} reveals the performance when the number of resources are varied and $K=15$. The results show that probabilistic feedback outperforms. And the performance of probabilistic feedback improves with the number of resource blocks $(N)$. However, in the case of error-bin, $\Upsilon^\texttt{FB}$ does not vary with the increase in number of $N$, though it remains low for lower values of $\wp$. Besides, we can see that optimal value of $\wp$ can be selected for probabilistic feedback depending on the available number of resources. For instance, when $N=15$ then $\wp=0.10$ is the optimal choice. In contrast, deterministic feedback is once again showing the worst performance.         

\begin{figure}
\centering
\includegraphics[scale=0.39]{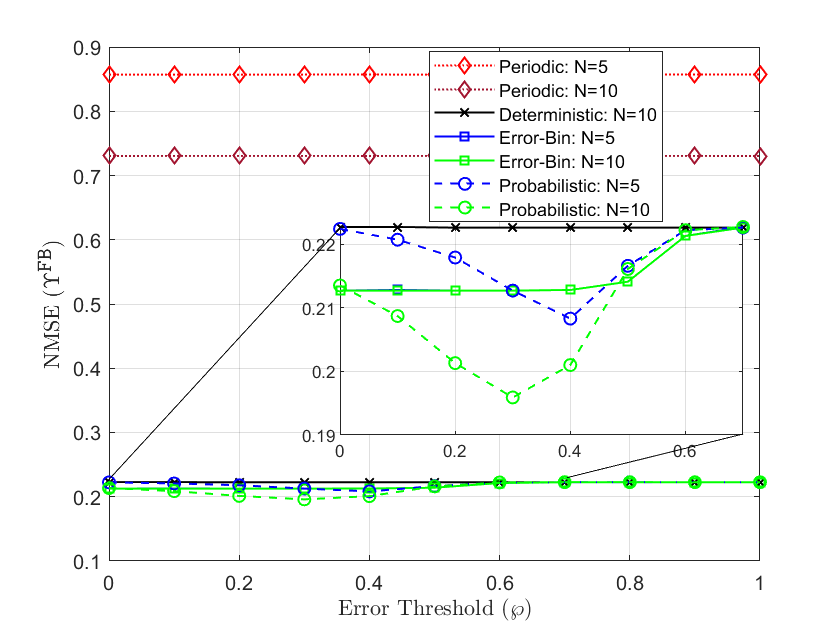}
\caption{A comparison of all CSI feedback schemes with different number of resource blocks $(N)$. $K=30$.}
\label{All_NMSE_FB_Ss_K30}
\end{figure}
Fig.\,\ref{All_NMSE_FB_Ss_K30} shows the performance when the UEs are increased to $K=30$ and when the performance is compared with periodic feedback. The results show that periodic feedback performs the worst despite reporting CSI to BS at regular intervals. However, periodic feedback improves when there are more resources as frequency of feedback increases, but still giving bad performing even when compared with deterministic feedback. The main reason behind this is the absence of CSILaBS as explained in Section\,\ref{periodic}. In contrast, probabilistic feedback outperforms rest of the schemes due to intelligent feedback selection. For instance, probabilistic feedback shows approximately $12\%$ improvement in comparison to deterministic feedback and $73\%$ compared to periodic when $\wp= 0.30$ and $N=10$. 
\subsection{Comparison with State-of-the-Art}
In this section, we compare the performance of CSILaBS with state-of-the-art, i.e., \textit{autoencoder} \cite{DL_CSIFB}, named CsiNet in \cite{DL_CSIFB}. For a fair comparison, we utilized pre-trained model of CsiNet, given in \cite{MATLAB}. Accordingly, we utilized the same dataset, as given in \cite{MATLAB}, to train CSILaBS. Briefly, the dataset, composed of $15$k samples, is generated using Matlab 5G toolbox function \textit{nrCDLChannel}. To get further details of the dataset and pre-trained CsiNet, please refer to \cite{MATLAB}.  Numerically, we observed a gain in $\varrho^\texttt{FB}$ of approximately $43\%$ when feedback CSI is highly compressed, i.e., $2$ bits are used to feedback CSI. Similarly, we observed a gain of around $5.5\,\text{dB}$ in $\Upsilon^\texttt{FB}$ for $2$ bits. Furthermore, increase in overhead bits resulted in less gain, which verifies our remark from Section\,\ref{CSILaBS} that increase in overhead brings no advantage in CSILaBS as there are no compression errors. However, in the case of multiuser, we can still get some gain due to the use of the proposed feedback selection algorithms. 

In terms of complexity, the number of floating-point operations for single-step prediction of CSILaBS are $5.8\,$M, which are very few in comparison to CsiNet ($58.52\,$M) \cite{CSI_FB_Survey}. Furthermore, CsiNet is composed of $32$ layers, and CSILaBS is of $4$ layers. This shows that a bigger
NN is required to design CsiNet and correspondingly resources to train it.        
\color{black}
\section{Conclusion}\label{Conclusions}
This paper investigated how ML can be avoided at UE for CSI feedback. Conventionally, ML has been considered at BS and UE, which makes it hard to train computationally expensive ML algorithm(s) at UE. In contrast, we addressed a novel method, coined CSILaBS, to efficiently recover compressed CSI without implementing ML at the UE. When applied ML at the BS, the generated function raised the idea of developing a codebook for downlink overhead reduction, which has shown highly standard-relevant implementation. Motivated by exploiting channel prediction for CSI feedback, we developed different ML-assisted CPs, which showed high performance gains compared to advanced ML models.

The results showed the effectiveness of CSILaBS compared to benchmark schemes. Numerical evaluations demonstrated an increase of approximately $22\%$ in precoding gain than without ML feedback information is highly compressed. Similarly, an approximately $65\%$ decrease in NMSE is observed for long-range uncompressed CSI prediction when the performance of hybrid model is compared to the advance time-series model, i.e., BiLSTM. 

Generally speaking, higher gain in the acquired channel at BS is achieved by using a minimum amount of overhead bits, that is, $2$. In contrast, the gain is negligible when $\geq5$ overhead bits are considered. The rationale behind nearly zero gain for $\geq5$ is minimum information loss while compressing the CSI. Importantly, the achieved gain came at without implementing ML on the UE-side. Thus, CSILaBS is the best choice when there is a high compression error, and UE is incapable for ML training. The results also verified that BiLSTM is beneficial when long-range CSI prediction is required on a massively compressed CSI, i.e., $1$ and $3$ quantization bits. In contrast, the hybrid model is the best option for low-compressed CSI prediction. Numerically, hybrid model has shown error of nearly $-21\,$dB and BiLSTM approximately $-15\,$dB for $\infty$ and $1$ quantization bits, respectively, while prediction is done for $10$ future indices.        
In the case of multiuser scenario, we observed that CSILaBS with the proposed probabilistic feedback method brings a gain of approximately $37\%$ in comparison to deterministic feedback, when $K=15$ and $N=10$ are considered. Furthermore, the gain drops to almost half when number of UEs are doubled, i.e., $K=30$. In a nutshell, we have learned that CSILaBS can retain ML gains without implementing ML at UE. CSILaBS can also work without using ML in the network, e.g., a simple Kalman filter can be exploited when the underlying channel model is evolving in a specific pattern.     
\ifCLASSOPTIONcaptionsoff
  \newpage
\fi



\bibliographystyle{IEEEtran}
\bibliography{mybib}

\begin{thebibliography}{10}
\providecommand{\url}[1]{#1}
\csname url@rmstyle\endcsname
\providecommand{\newblock}{\relax}
\providecommand{\bibinfo}[2]{#2}
\providecommand\BIBentrySTDinterwordspacing{\spaceskip=0pt\relax}
\providecommand\BIBentryALTinterwordstretchfactor{4}
\providecommand\BIBentryALTinterwordspacing{\spaceskip=\fontdimen2\font plus
\BIBentryALTinterwordstretchfactor\fontdimen3\font minus \fontdimen4\font\relax}
\providecommand\BIBforeignlanguage[2]{{%
\expandafter\ifx\csname l@#1\endcsname\relax
\typeout{** WARNING: IEEEtran.bst: No hyphenation pattern has been}%
\typeout{** loaded for the language `#1'. Using the pattern for}%
\typeout{** the default language instead.}%
\else
\language=\csname l@#1\endcsname
\fi
#2}}

\bibitem{karam_CP_Globecom}
M.~K. Shehzad, L.~Rose, M.~F. Azam, and M.~Assaad, ``{Real-time massive MIMO channel prediction: A combination of deep learning and NeuralProphet},'' in \emph{GLOBECOM 2022 - 2022 IEEE Global Communications Conference}, 2022, pp. 1423--1428.

\bibitem{Karam_AI}
M.~K. Shehzad, L.~Rose, M.~M. Butt, I.~Z. Kovacs, M.~Assaad, and M.~Guizani, ``{Artificial intelligence for 6G networks: Technology advancement and standardization},'' \emph{IEEE Vehicular Technology Magazine}, vol.~17, no.~3, pp. 16--25, 2022.

\bibitem{CSI_TCCN_Karam}
\BIBentryALTinterwordspacing
M.~K. Shehzad, L.~Rose, S.~Wesemann, M.~Assaad, and S.~A. Hassan, ``{Design of an efficient CSI feedback mechanism in massive MIMO systems: A machine learning approach using empirical data},'' 2022. [Online]. Available: \url{https://arxiv.org/abs/2208.11951}
\BIBentrySTDinterwordspacing

\bibitem{AI_Real_World}
J.~Guo, X.~Li, M.~Chen, P.~Jiang, T.~Yang, W.~Duan, H.~Wang, S.~Jin, and Q.~Yu, ``{AI enabled wireless communications with real channel measurements: Channel feedback},'' \emph{Journal of Communications and Information Networks}, vol.~5, no.~3, pp. 310--317, 2020.

\bibitem{Karam_CP}
M.~K. Shehzad, L.~Rose, S.~Wesemann, and M.~Assaad, ``{ML-based massive MIMO channel prediction: Does it work on real-world data?}'' \emph{IEEE Wireless Communications Letters}, vol.~11, no.~4, pp. 811--815, 2022.

\bibitem{CS_FB_precoding_arxiv}
F.~Carpi, S.~Venkatesan, J.~Du, H.~Viswanathan, S.~Garg, and E.~Erkip, ``{Precoding-oriented massive MIMO CSI feedback design},'' \emph{arXiv preprint arXiv:2302.11526}, 2023.

\bibitem{Karam_Dealing}
M.~K. Shehzad, L.~Rose, and M.~Assaad, ``{Dealing with CSI compression to reduce losses and overhead: An artificial intelligence approach},'' in \emph{2021 IEEE International Conference on Communications Workshops (ICC Workshops)}, 2021, pp. 1--6.

\bibitem{DL_CSIFB}
C.-K. Wen, W.-T. Shih, and S.~Jin, ``{Deep learning for massive MIMO CSI feedback},'' \emph{IEEE Wireless Communications Letters}, vol.~7, no.~5, pp. 748--751, 2018.

\bibitem{Karam_UAV}
M.~K. Shehzad, L.~Rose, and M.~Assaad, ``{RNN-based twin channel predictors for CSI acquisition in UAV-assisted 5G+ networks},'' in \emph{2021 IEEE Global Communications Conference (GLOBECOM)}, 2021, pp. 1--6.

\bibitem{TWC_Ref1}
H.~Hojatian, J.~Nadal, J.-F. Frigon, and F.~Leduc-Primeau, ``{Unsupervised deep learning for massive MIMO hybrid beamforming},'' \emph{IEEE Transactions on Wireless Communications}, vol.~20, no.~11, pp. 7086--7099, 2021.

\bibitem{CSI_FB_Rana}
R.~Ahmed, K.~Jayasinghe, and T.~Wild, ``{Comparison of explicit CSI feedback schemes for 5G new radio},'' in \emph{2019 IEEE 89th Vehicular Technology Conference (VTC2019-Spring)}, 2019, pp. 1--5.

\bibitem{CSI_FB_Survey}
J.~Guo, C.-K. Wen, S.~Jin, and G.~Y. Li, ``{Overview of deep learning-based CSI feedback in massive MIMO systems},'' \emph{IEEE Transactions on Communications}, vol.~70, no.~12, pp. 8017--8045, 2022.

\bibitem{CSI_FB_Reciprocity}
E.~Becirovic, E.~Björnson, and E.~G. Larsson, ``{Combining reciprocity and CSI feedback in MIMO systems},'' \emph{IEEE Transactions on Wireless Communications}, vol.~21, no.~11, pp. 10\,065--10\,080, 2022.

\bibitem{CSI_FB_Nokia}
M.~Kurras, S.~Jaeckel, L.~Thiele, and V.~Braun, ``{CSI compression and feedback for network MIMO},'' in \emph{2015 IEEE 81st Vehicular Technology Conference (VTC Spring)}, 2015, pp. 1--5.

\bibitem{3GPP_5G}
H.~Holma, A.~Toskala, and T.~Nakamura, \emph{{5G technology: 3GPP new radio}}.\hskip 1em plus 0.5em minus 0.4em\relax John Wiley \& Sons, 2020.

\bibitem{CSI_FB_TWC1}
Z.~Hu, G.~Liu, Q.~Xie, J.~Xue, D.~Meng, and D.~G{\"u}nd{\"u}z, ``{A learnable optimization and regularization approach to massive MIMO CSI feedback},'' \emph{IEEE Transactions on Wireless Communications}, 2023.

\bibitem{CSI_FB_TWC2}
M.~Nerini, V.~Rizzello, M.~Joham, W.~Utschick, and B.~Clerckx, ``{Machine learning-based CSI feedback with variable length in FDD massive MIMO},'' \emph{IEEE Transactions on Wireless Communications}, vol.~22, no.~5, pp. 2886--2900, 2023.

\bibitem{CSIFB_new_1}
J.~Guo, L.~Wang, F.~Li, and J.~Xue, ``{CSI feedback with model-driven deep learning of massive MIMO systems},'' \emph{IEEE Communications Letters}, vol.~26, no.~3, pp. 547--551, 2022.

\bibitem{CSIFB_new_2}
S.~Tang, J.~Xia, L.~Fan, X.~Lei, W.~Xu, and A.~Nallanathan, ``{Dilated convolution based CSI feedback compression for massive MIMO systems},'' \emph{IEEE Transactions on Vehicular Technology}, pp. 1--6, 2022.

\bibitem{CSI_FB_TWC3}
J.~Guo, C.-K. Wen, S.~Jin, and G.~Y. Li, ``{Convolutional neural network-based multiple-rate compressive sensing for massive MIMO CSI feedback: Design, simulation, and analysis},'' \emph{IEEE Transactions on Wireless Communications}, vol.~19, no.~4, pp. 2827--2840, 2020.

\bibitem{CSIFB_new_3}
H.~Li, B.~Zhang, H.~Chang, X.~Liang, and X.~Gu, ``{CVLNet: A complex-valued lightweight network for CSI feedback},'' \emph{IEEE Wireless Communications Letters}, vol.~11, no.~5, pp. 1092--1096, 2022.

\bibitem{CSI_FB_TWC4}
X.~Liang, H.~Chang, H.~Li, X.~Gu, and L.~Zhang, ``{Changeable rate and novel quantization for CSI feedback based on deep learning},'' \emph{IEEE Transactions on Wireless Communications}, vol.~21, no.~12, pp. 10\,100--10\,114, 2022.

\bibitem{CSIFB_new_5}
G.~Fan, J.~Sun, G.~Gui, H.~Gacanin, B.~Adebisi, and T.~Ohtsuki, ``{Fully convolutional neural network-based CSI limited feedback for FDD massive MIMO systems},'' \emph{IEEE Transactions on Cognitive Communications and Networking}, vol.~8, no.~2, pp. 672--682, 2022.

\bibitem{CSI_FB_TWC5}
M.~B. Mashhadi, Q.~Yang, and D.~Gündüz, ``{Distributed deep convolutional compression for massive MIMO CSI feedback},'' \emph{IEEE Transactions on Wireless Communications}, vol.~20, no.~4, pp. 2621--2633, 2021.

\bibitem{Auto_Encoder_1}
Y.~Jang, G.~Kong, M.~Jung, S.~Choi, and I.-M. Kim, ``{Deep autoencoder based CSI feedback with feedback errors and feedback delay in FDD massive MIMO systems},'' \emph{IEEE Wireless Communications Letters}, vol.~8, no.~3, pp. 833--836, 2019.

\bibitem{IEEE_WCL1}
Z.~Lu, X.~Zhang, R.~Zeng, and J.~Wang, ``{Better lightweight network for free: Codeword mimic learning for massive MIMO CSI feedback},'' \emph{IEEE Communications Letters}, 2023.

\bibitem{IEEE_WCL2}
S.~Mourya, S.~Amuru, and K.~K. Kuchi, ``{A spatially separable attention mechanism for massive MIMO CSI feedback},'' \emph{IEEE Wireless Communications Letters}, vol.~12, no.~1, pp. 40--44, 2023.

\bibitem{IEEE_WCL3}
B.~Zhang, H.~Li, X.~Liang, X.~Gu, and L.~Zhang, ``{Multi-task training approach for CSI feedback in massive MIMO systems},'' \emph{IEEE Communications Letters}, vol.~27, no.~1, pp. 200--204, 2023.

\bibitem{Auto_Encoder_2}
T.~Wang, C.-K. Wen, S.~Jin, and G.~Y. Li, ``{Deep learning-based CSI feedback approach for time-varying massive MIMO channels},'' \emph{IEEE Wireless Communications Letters}, vol.~8, no.~2, pp. 416--419, 2019.

\bibitem{Auto_Encoder_3}
C.~Lu, W.~Xu, H.~Shen, J.~Zhu, and K.~Wang, ``{MIMO channel information feedback using deep recurrent network},'' \emph{IEEE Communications Letters}, vol.~23, no.~1, pp. 188--191, 2019.

\bibitem{CS_FB_Arxiv}
W.~Chen, W.~Wan, S.~Wang, P.~Sun, and B.~Ai, ``{CSI-PPPNet: A one-sided deep learning framework for Massive MIMO CSI feedback},'' \emph{arXiv preprint arXiv:2211.15851}, 2022.

\bibitem{Auto_Encoder_4}
X.~Li and H.~Wu, ``{Spatio-temporal representation with deep neural recurrent network in MIMO CSI feedback},'' \emph{IEEE Wireless Communications Letters}, vol.~9, no.~5, pp. 653--657, 2020.

\bibitem{Auto_Encoder_5}
Z.~Liu, M.~del Rosario, and Z.~Ding, ``{A Markovian model-driven deep learning framework for massive MIMO CSI feedback},'' \emph{IEEE Transactions on Wireless Communications}, vol.~21, no.~2, pp. 1214--1228, 2022.

\bibitem{CSI_FB_TWC6}
Z.~Lu, X.~Zhang, H.~He, J.~Wang, and J.~Song, ``{Binarized aggregated network with quantization: Flexible deep learning deployment for CSI feedback in massive MIMO systems},'' \emph{IEEE Transactions on Wireless Communications}, vol.~21, no.~7, pp. 5514--5525, 2022.

\bibitem{Autoencoder_Disadvantages}
\BIBentryALTinterwordspacing
D.~Bank, N.~Koenigstein, and R.~Giryes, ``Autoencoders,'' 2020. [Online]. Available: \url{https://arxiv.org/abs/2003.05991}
\BIBentrySTDinterwordspacing

\bibitem{Karam_KF}
M.~K. Shehzad, L.~Rose, and M.~Assaad, ``{A novel algorithm to report CSI in MIMO-based wireless networks},'' in \emph{ICC 2021 - IEEE International Conference on Communications}, 2021, pp. 1--6.

\bibitem{DL_channel_estimation}
M.~Soltani, V.~Pourahmadi, A.~Mirzaei, and H.~Sheikhzadeh, ``Deep learning-based channel estimation,'' \emph{IEEE Communications Letters}, vol.~23, no.~4, pp. 652--655, 2019.

\bibitem{CP_age}
J.~Yuan, H.~Q. Ngo, and M.~Matthaiou, ``{Machine learning-based channel prediction in massive MIMO with channel aging},'' \emph{IEEE Transactions on Wireless Communications}, vol.~19, no.~5, pp. 2960--2973, 2020.

\bibitem{CP_TWC1}
H.~Kim, J.~Choi, and D.~J. Love, ``{Massive MIMO channel prediction via meta-learning and deep denoising: Is a small dataset enough?}'' \emph{IEEE Transactions on Wireless Communications}, pp. 1--1, 2023.

\bibitem{CP_TWC2}
M.~Chu, A.~Liu, V.~K.~N. Lau, C.~Jiang, and T.~Yang, ``{Deep reinforcement learning based end-to-end multiuser channel prediction and beamforming},'' \emph{IEEE Transactions on Wireless Communications}, vol.~21, no.~12, pp. 10\,271--10\,285, 2022.

\bibitem{CP_TWC3}
X.~Wang, Y.~Shi, W.~Xin, T.~Wang, G.~Yang, and Z.~Jiang, ``{Channel prediction with time-varying Doppler spectrum in high-mobility scenarios: A polynomial Fourier transform based approach and field measurements},'' \emph{IEEE Transactions on Wireless Communications}, pp. 1--1, 2023.

\bibitem{CP_TWC4}
F.~Peng, S.~Zhang, Z.~Jiang, X.~Wang, and W.~Chen, ``{A novel mobility induced channel prediction mechanism for vehicular communications},'' \emph{IEEE Transactions on Wireless Communications}, vol.~22, no.~5, pp. 3488--3502, 2023.

\bibitem{CP_TWC5}
W.~Li, H.~Yin, Z.~Qin, Y.~Cao, and M.~Debbah, ``{A multi-dimensional matrix pencil-based channel prediction method for massive MIMO with mobility},'' \emph{IEEE Transactions on Wireless Communications}, vol.~22, no.~4, pp. 2215--2230, 2023.

\bibitem{CP_TWC6}
T.~Zhou, H.~Zhang, B.~Ai, C.~Xue, and L.~Liu, ``{Deep-learning-based spatial–temporal channel prediction for smart high-speed railway communication networks},'' \emph{IEEE Transactions on Wireless Communications}, vol.~21, no.~7, pp. 5333--5345, 2022.

\bibitem{neuralprophet}
O.~Triebe, H.~Hewamalage, P.~Pilyugina, N.~Laptev, C.~Bergmeir, and R.~Rajagopal, ``{NeuralProphet: Explainable forecasting at scale},'' \emph{arXiv preprint arXiv:2111.15397}, 2021.

\bibitem{Grid_search}
\BIBentryALTinterwordspacing
P.~Liashchynskyi and P.~Liashchynskyi, ``{Grid search, random search, genetic algorithm: A big comparison for NAS},'' 2019. [Online]. Available: \url{https://arxiv.org/abs/1912.06059}
\BIBentrySTDinterwordspacing

\bibitem{AR-Net}
O.~Triebe, N.~Laptev, and R.~Rajagopal, ``{\textit{AR-Net}: A simple auto-regressive neural network for time-series},'' \emph{arXiv preprint arXiv:1911.12436}, 2019.

\bibitem{data_augmentation}
X.~Liang, Z.~Liu, H.~Chang, and L.~Zhang, ``{Wireless channel data augmentation for artificial intelligence of things in industrial environment using generative adversarial networks},'' in \emph{2020 IEEE 18th International Conference on Industrial Informatics (INDIN)}, vol.~1, 2020, pp. 502--507.

\bibitem{CSMA_1}
J.~Ghaderi and R.~Srikant, ``{On the design of efficient CSMA algorithms for wireless networks},'' in \emph{49th IEEE Conference on Decision and Control (CDC)}.\hskip 1em plus 0.5em minus 0.4em\relax IEEE, 2010, pp. 954--959.

\bibitem{CSMA_2}
J.~Ni, B.~Tan, and R.~Srikant, ``{Q-CSMA: Queue-length-based CSMA/CA algorithms for achieving maximum throughput and low delay in wireless networks},'' \emph{IEEE/ACM Transactions on Networking}, vol.~20, no.~3, pp. 825--836, 2011.

\bibitem{MATLAB}
``{CSI feedback with autoencoders},'' \url{https://de.mathworks.com/help/comm/ug/csi-compression-autoencoder.html#mw_rtc_CSICompressionAutoencoderExample_M_A50E54F6}, accessed on 2023-06-10. [online].

\end{thebibliography}
\begin{IEEEbiography}[{\includegraphics[width=1.3in,height=1.25in,clip,keepaspectratio]{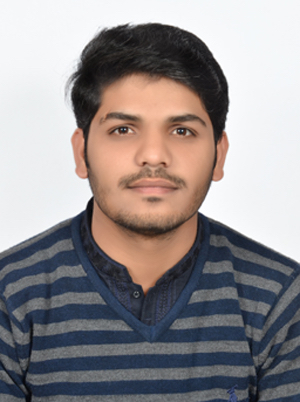}}]{M. Karam Shehzad} (Member, IEEE) is a senior research specialist with Nokia Standards, Paris, France. He received his Ph.D. in Networks, Information and Communication Sciences from CentraleSupélec, University of Paris-Saclay, Paris, France, in 2023. He received M.S. in Electrical Engineering from National University of Sciences and Technology (NUST), Islamabad, Pakistan, in 2019. During his M.S., he also spent one semester on ERASMUS+ mobility program at University of Malaga, Malaga, Spain. He received his BEng. (Hons) in Electrical and Electronic Engineering from University of Bradford, Bradford, UK, in 2016. From November 2019 to February 2020, he was a research assistant at NUST, Islamabad, Pakistan. From 2016 to 2017, he worked as a research assistant at Namal University, Mianwali, Pakistan. His research and standardization interests include MIMO communication and AI/ML for wireless communications. 
\end{IEEEbiography}
\vskip -2\baselineskip plus -1fil
\begin{IEEEbiography}[{\includegraphics[width=1.0in,height=1.25in,clip,keepaspectratio]{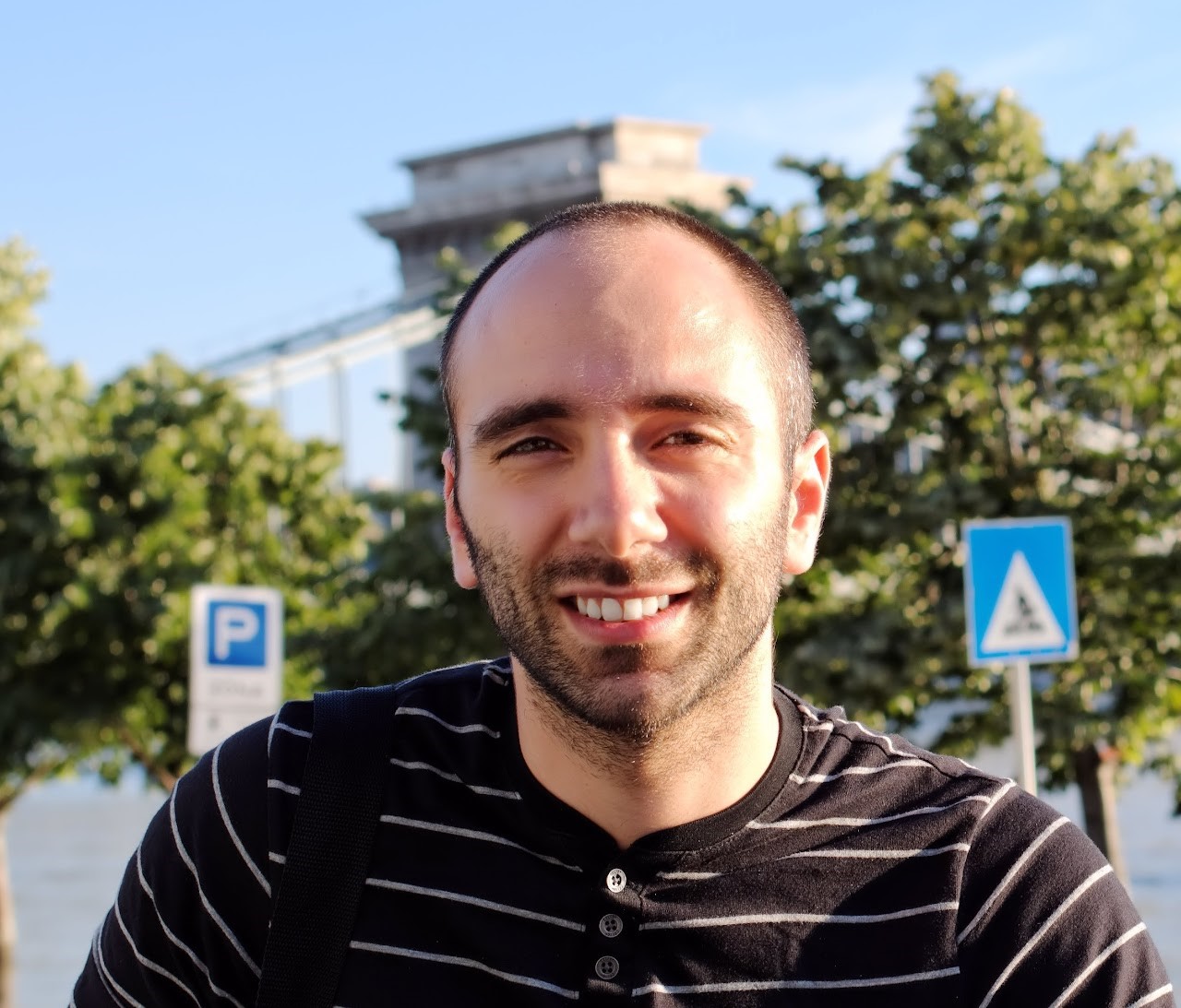}}]{Luca Rose} (Member, IEEE) is a senior research and standardization expert with Nokia Standards, Paris, France. He received his M.Sc. from the University of Pisa, Italy, and his Ph.D. in physics from Centrale-Supelec, Gif-sur-Yvette, France, in 2013. He worked at Huawei France research center, Sigfox, and Thales Communications and Security, implementing the first IoT dedicated network, studying MIMO-related problems, and contributing to several standards. He is currently an ITU-R and ETSI delegate investigating telecommunication coexistence issues and dynamic spectrum allocation. His interests span from the field of AI-ML applied to channel acquisition, to physical layer security and game theory. Furthermore, he is Lead Editor of the Internet of Things Series of IEEE Communications Magazine.
\end{IEEEbiography}
\vskip -2\baselineskip plus -1fil
\begin{IEEEbiography}[{\includegraphics[width=1.3in,height=1.25in,clip,keepaspectratio]{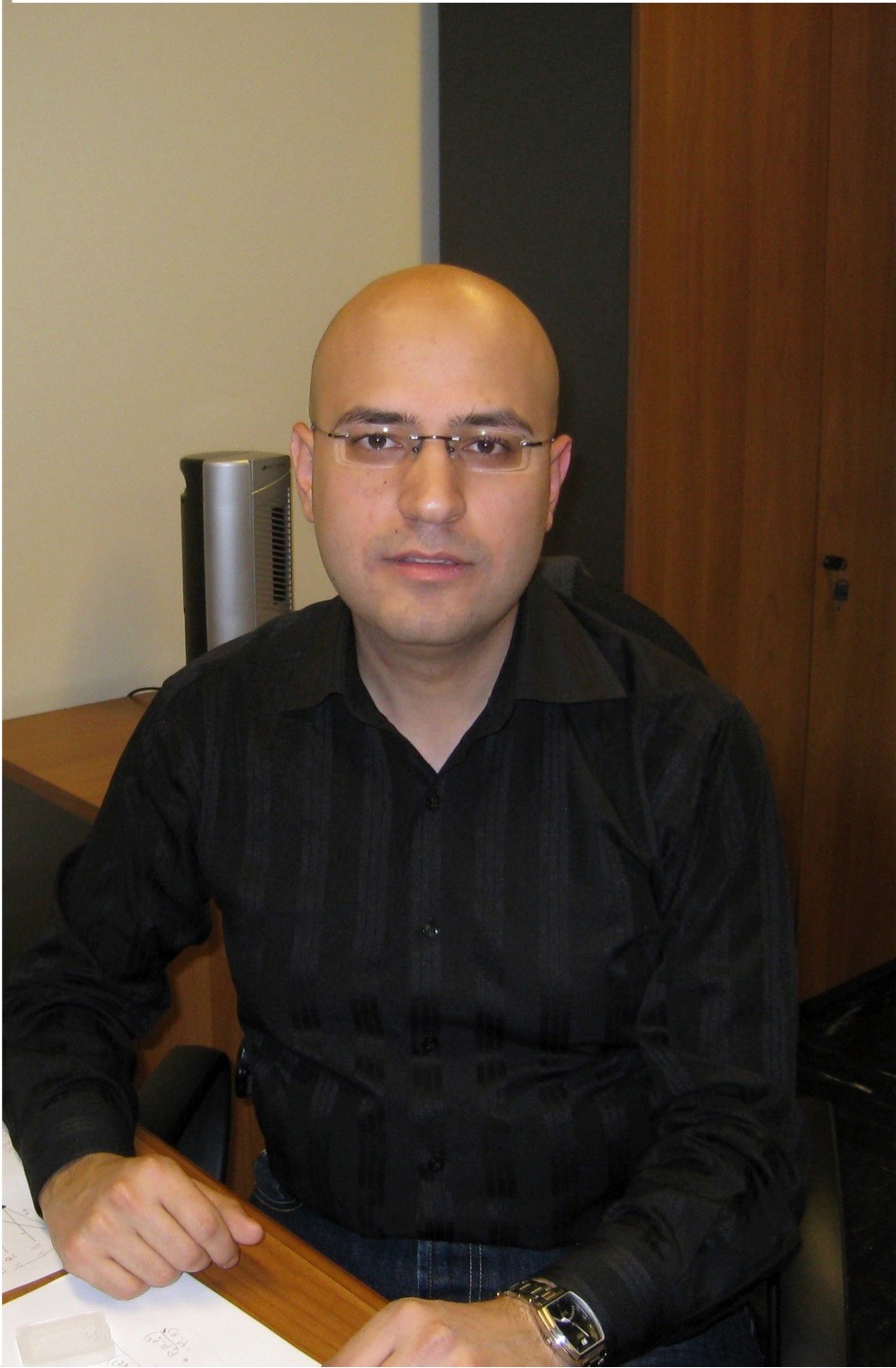}}]{Mohamad Assaad} (Senior Member, IEEE) received the M.Sc. and Ph.D. degrees in telecommunications from Telecom ParisTech, Paris, France, in 2002 and 2006, respectively. Since 2006, he has been with the Telecommunications Department, CentraleSupelec, University of Paris-Saclay, where he is currently a professor. He is also a Researcher with the Laboratoire des Signaux et systèmes (CNRS) and was the 5G Chair Holder between 2017 and 2021. He has coauthored one book and more than 150 publications in journals and conference proceedings. He has served as TPC member or TPC co-chair for various top-tier international conferences, including TPC co-chair for IEEE WCNC'21, IEEE Globecom'20 Mobile and Wireless networks Symposium Co-chair, etc. He has served also as Editor for several international journals, including editor for the IEEE Wireless Communications Letters between 2017 and 2022, Guest Editor for the IEEE Transactions on Network Science and Engineering, etc. He has also given in the past successful tutorials on topics related to 5G systems, and Age of Information at various IEEE conferences. His research interests include 5G and 6G systems, fundamental networking aspects of wireless systems, Age of Information and semantic communications, resource optimization and machine learning in wireless networks.
\end{IEEEbiography}







\end{document}